\documentclass[aps,prl,showpacs,twocolumn, groupedaddress,amssymb,amsmath]{revtex4-1}
\usepackage{color, graphicx}
\usepackage{amsmath}
\usepackage{dcolumn}
\usepackage{bm}

\begin{document}

\title{Multi-component Transparent Conducting Oxides: Progress in Materials Modelling}

\author{Aron Walsh$^{1*}$, Juarez L. F. Da Silva$^2$ and Su-Huai Wei$^3$}
\affiliation{$^1$University College London, Kathleen Lonsdale Materials Chemistry, Department of Chemistry, 20 Gordon Street, London WC1H 0AJ, UK}
\affiliation{$^2$Instituto de Fisica de S\~ao Carlos, Universidade de S\~ao Paulo, Cx. Postal 369, S\~ao Carlos 13560-970, SP, Brazil}
\affiliation{$^3$National Renewable Energy Laboratory, Golden, CO 80401, USA}

\date{\today}


\begin{abstract}
Transparent conducting oxides (TCOs) play an essential role in modern optoelectronic devices through their combination of electrical conductivity and optical transparency. We review recent progress in our understanding of multi-component TCOs formed from solid-solutions of ZnO, In$_2$O$_3$, Ga$_2$O$_3$ and Al$_2$O$_3$, with a particular emphasis on the contributions of materials modelling, primarily based on Density Functional Theory.  In particular, we highlight three major results from our work: (i) the fundamental principles governing the crystal structures of multi-component oxide structures including (In$_2$O$_3$)(ZnO)$_n$, named IZO, and (In$_2$O$_3$)$_m$(Ga$_2$O$_3$)$_l$(ZnO)$_n$, named IGZO; (ii) the relationship between elemental composition and optical and electrical behaviour, including valence band alignments; (iii) the high-performance of amorphous oxide semiconductors. From these advances, the challenge of the rational design of novel electroceramic materials is discussed.
\end{abstract}

\maketitle

\section{Introduction}
Transparent conducting oxides (TCOs) are defined by high electrical conductivity approaching that of a metallic compound,
with the high transmission of photons in the visible range of the electromagnetic spectrum.
The unique combination of these two features makes TCOs essential components of modern optoelectronic devices\cite{thomas-907, edwards-2295}.
The common electronic characteristic of $n$-type TCO materials is a spatially delocalised, low energy, and low effective mass conduction band determined
primarily by the metal $s$ orbitals. Such a state is achieved with oxides formed from closed-shell $ns^0$ cations such as Zn(II), Ga(III) and In(III).

The prototype $n$-type TCO is Sn-doped In$_2$O$_3$ (ITO), which has been the subject of intense study both experimentally\cite{ohya-240, wit-143, wit-142, weiher-2834, edwards-2295, hamberg-3240, king-116808, korber-165207, klein-1197} and theoretically\cite{walsh-167402, walsh-075211, walsh-10438, lany-045501, agoston-455801, tomita-051911, agoston-245501, medvedeva-125116}.
Other examples of electron conducting TCOs include Al-doped ZnO, Ga-doped ZnO and F-doped SnO$_2$. Typically these materials have electron carrier concentrations ranging from 10$^{16}$ - 10$^{21}$ cm$^{-3}$, and electron mobility ranging from 10 to 1000 cm$^2$V$^{-1}$s$^{-1}$, depending on the material quality and growth conditions\cite{edwards-2295}.

In contrast to the delocalised conduction band wave-functions, the valence band states of hetero-polar metal oxides are typically localised O 2$p$ states. Instead of resulting in mobile electron holes, acceptor doping of these materials results in deep localised states, which are stabilised by local lattice distortions - the formation of small polarons\cite{schirmer-667, stoneham-255208}. To obtain hole conducting ($p$-type) TCO materials, the principal approach has been to include a metal with filled low binding energy states at the top of the valence band, such as Cu(I), which can facilitate hole transport\cite{nie-066405, nie-075111}.
Following the work of Hosono \textit{et al.} on CuAlO$_2$\cite{kawazoe-939}, many ternary Cu oxides have been explored for this purpose. The limitation of this class of material is the poor hole mobility ($<$ 10 cm$^2$V$^{-1}$s$^{-1}$) that arises from the relatively localised Cu $d$ bands, as well as the indirect electronic band gaps arising from the underlying lattice symmetry. Recent work has focused on other delafossite (Cu$M^{III}$O$_2$, where $M^{III}$ is a trivalent cation) materials including CuGaO$_2$\cite{nie-066405}, CuCrO$_2$\cite{saadi-272, scanlon-035101, arnold-075102}, CuScO$_2$\cite{huda-035205} and CuBO$_2$\cite{scanlon-4568}, in addition to studies on related ternary copper oxides containing Sr and Pb\cite{kudo-220, nie-075111, godinho-2798}. Unfortunately for all Cu-based materials, the poor carrier mobility, which is directly related to the parent compound Cu$_2$O, is difficult to overcome. In some cases, the existence of deep defect centers\cite{scanlon-096405}, which act as trapping centres for free carriers, pin the Fermi level well above the valence band, thus limiting the carrier densities achievable by doping.

The possibility of discovering alternative $n$-type TCO compositions with superior material properties has led to the investigation of ternary and quaternary systems, with In$_2$O$_3$(ZnO)$_n$ (IZO) being one particular focus of attention\cite{minami-971, taylor-90, hiramatsu-3033, kumar-073703,leenheer-115215, moriga-312}. These studies have highlighted the improved chemical and thermal stability of IZO compared to ITO, making it more desirable for commercial application. One issue with the utilisation of multi-component oxide materials is the difficulty in synthesising high-quality crystalline samples at low cost. To overcome this limitation, there has been an active interest in growing amorphous TCOs (a-TCOs)\cite{nomura-488, sun-1897, lee-843, taylor-3169, kumar-073703}. Amorphous oxides can exhibit electrical properties comparable to their crystalline phases, in contrast to traditional amorphous semiconductors. In amorphous systems, the underlying crystalline nature and translational symmetry of a material is lost; therefore, many atoms have unsaturated dangling bonds. For covalent materials such as Si, the dangling bond states are located at the center of the band gap, acting as both electron and hole traps, therefore, amorphisation of Si is associated with a substantial increase in electrical resistance, relative to the crystalline phase\cite{mott-1987}. For ionic or polar oxides, the dangling bond states are very close to the band edges; therefore, amorphisation of oxides usually does not create deep trap levels, which leads to the remarkable fact that a-TCOs can exhibit comparable transport properties relative to their crystalline counterparts, an effect that has only recently been understood through materials modelling\cite{walsh-5119, robertson-1026, medvedeva-125116, hosono-2796}.

The physical and chemical properties of TCO materials have been the subject of many excellent reviews. In particular, the pioneering experimental work of Hosono \textit{et al.} has been recently reviewed\cite{hosono-6000}, while a more comprehensive review of the field can be found in a themed issue of the Material Research Bulletin\cite{fortunato-242}. The chemistry of the ZnO-In$_2$O$_3$-SnO$_2$ system was the focus of a detailed review by Hoel \textit{et al.}\cite{hoel-3569}. From a theoretical perspective, comprehensive overviews focused on ZnO can be found in the works of Catlow \textit{et al.}\cite{catlow-2234}, Lany and Zunger\cite{lany-235104} and Janotti and Van de Walle\cite{janotti-165202}. In addition, Medvedeva and Hettiarachchi recently provided an in-depth analysis of the properties of complex TCOs based on trends in their physical properties\cite{medvedeva-125116}.

In this Topical Review, we focus on recent advances in the theoretical understanding of multi-component TCOs.
After briefly addressing the binary metal oxide components of the complex TCOs, we address a number of highlights from our recent investigations, including:
(i) the fundamental principles governing the structures of multi-component crystalline and amorphous oxide systems;
(ii) the relationship between composition and optoelectronic behaviour;
(iii) the origin of high-performance for amorphous oxide semiconductors.
From these advances, future directions in the field of transparent conducting semiconductors are discussed.

\section{Computational Methods}

The workhorse technique in applied computational material science is Density
Functional Theory (DFT)\cite{dft1,dft2}, for which Walter Kohn and John Pople shared the
Nobel prize in 1998. In contrast to quantum chemical methods, which are
concerned with the $3N$ dimensional many-body wave-function, in DFT it is the
electron density that is the key quantity to be optimised and from which all
ground-state properties can be determined. The result is an {\it ab initio}
method that can be applied to realistic material simulations in solid-state
physics, organic chemistry, soft matter and related fields.

Approaches based on DFT are widely available in a number of academic and commercial
packages, \textit{e.g.} the codes
\textit{VASP}\cite{vasp1}, \textit{CASTEP}\cite{castep}, \textit{Quantum-Espresso}\cite{espresso} and \textit{FHI-AIMS}\cite{aims1}.
The principal difference between all implementations is in the
choice of the basis set, which is used to represent the single-particle
Kohn-Sham wave-functions, $e.g.$ plane-waves, augmented plane-waves with atomic
orbitals, gaussian functions, Slater orbitals, numerical orbitals.
The technical details of these approaches have been reviewed in detail
elsewhere\cite{martin-2004, payne-1045}.

Evaluation of the total energy, within the DFT formalism, gives access to a
range of thermodynamic properties including heats of formation, ionisation
energies and phase stability. Furthermore, forces and stress can be
calculated routinely, which provides the flexibility for a wide range of
studies on structure determination, including, for example,
global optimisation\cite{woodley-937, walsh-8446, trimarchi-295212, hautier-3762}.
The solution of the Kohn-Sham equations
yield the eigenvectors and eigenvalues for the respective system, and hence, it
can be used to understand the underlying bonding mechanics and electrical
transport properties.

The limitations of DFT are concentrated in the exchange-correlation
functional, which contains the description of all quantum mechanical effects
for the system of interest. The exact form of the functional remains unknown
and must be approximated, which is generally in the form of local or
semi-local functionals, \textit{i.e.} the local density approximation\cite{lda} (LDA)
or generalised gradient approximation\cite{pbe} (GGA), respectively. Both
approaches gives rise to an accurate description of the structural, electronic and
thermodynamic parameters for most systems; however, they fail to give a proper
description of particular properties and systems, in particular, highly correlated (strongly interacting)
electronic states\cite{perdew-5048}.

It has been well demonstrated that both the LDA and GGA exchange-correlation functionals
underestimate the electronic band gaps of insulators and semiconductors due to a
discontinuity in the wavefunction character at the band edges\cite{perdew-1884, sham-1888}.
For quantitative band gap predictions,
the use of many-body approaches such as $GW$ theory\cite{hedin-796}
are necessary; however, the use of non-local hybrid density functionals
offers a means to reduce the error within the single-particle formalism.
In the hybrid approach, a percentage of exact non-local Hartree-Fock exchange
is added to a local or semi-local exchange-correlation functional.
One caveat of this approach is that the amount of exact-exchange
and spatial screening remains material and property dependent\cite{cora-2004}.
One hybrid functional, which has had particular success in its
application to oxides, is the Heyd-Scuseria-Ernzerhof (HSE06) formulation\cite{hse, paier-154709},
in which 25\% of exact exchange replaces the short-range GGA exchange potential.

\section{Binary Oxides}

Here, we briefly discuss the bulk properties of the main binary components of
the multi-ternary systems of interest, \textit{i.e.} ZnO, In$_2$O$_3$, Ga$_2$O$_3$,
and Al$_2$O$_3$. A number of the material properties are summarised in
Table~\ref{t1}, including the band gaps,
heat of formation, cation effective coordination numbers, and the partial
charges resulting from Bader analysis\cite{bader,bader-vasp} of the
equilibrium electron density. Absolute values of the partial charges derived from
delocalised electron densities are not unique, and are generally smaller than
the formal ionic charge\cite{jansen-10026,catlow-4321}, but relative changes
in these values can still be instructive.

\subsection{ZnO}

ZnO is a material that has a rich solid-state physics and chemistry,
with a range of applications that exploit its electronic, optical and
piezoelectric properties\cite{ozgur-041301}. Recently, magnetic doping of ZnO
has been the subject of intense study, \textit{e.g.} Co doped ZnO, where the
ferromagnetic coupling between Co $d$ states can be electron mediated
\cite{kittilstved-291,walsh-256401, walsh-159702}.
In its ground-state, ZnO adopts the
hexagonal wurtzite (wz) crystal structure (space group
$P6_3mc$) with lattice constants of $a_0^{\rm exp} = 3.250$~{\AA} and $c_0^{\rm
exp} = 5.207$~{\AA},\cite{ozgur-041301}
in which the Zn and O ions form roughly tetrahedral
coordinated environments. The experimental electronic band gap
of ZnO is 3.44 eV\cite{madelung-04}, which is direct in nature. As mentioned in the
Introduction, the material properties of ZnO have been subject of recent
review both from an experimental\cite{ozgur-041301} and computational
perspective\cite{catlow-2234, janotti-165202, lany-235104, catlow-1923, sokol-267}.

\subsection{In$_2$O$_3$}

The high $n$-type conductivity of In$_2$O$_3$ is exploited in many TCO device
applications, which has greatly contributed to the rising cost of In metal\cite{hamberg-r123}.
Indium oxide adopts the body-centred cubic (bcc) bixbyite lattice (space
group $Ia\bar3$) with $a_0^{\rm exp} = 10.117$~{\AA}\cite{marezio-723},
which is a defective $2\times2\times2$ superstructure of the fluorite mineral structure.
The In cations form roughly octahedral
structures, which implies an effective coordination number close to 6. As
mentioned above, In$_2$O$_3$ is widely used as a TCO; however, the 
long-standing  band gap problem of In$_2$O$_3$ was only
recently resolved from a joint experimental and theoretical effort\cite{walsh-167402}.
The direct electronic band gap of the bulk material is of the order of 2.9 eV\cite{walsh-167402},
and the large disparity between the electronic and optical band gaps
($E_g^{opt} \sim 3.6$~eV) arises from a combination
of dipole forbidden optical transitions
and conduction band occupation\cite{walsh-167402, walsh-075211, korber-165207, fuchs-155107}.

\subsection{Ga$_2$O$_3$}

Ga$_2$O$_3$ is a wide band gap metal oxide ($E_g > 4.5$~eV), which is used in
lasers, as well as a dielectric coating in electrical devices\cite{passlack-686}.
 Of the five crystal polymorphs, the stable $\beta$ phase (space
group $C2/m$) consists of close packed oxide ions, with Ga occupying a
combination of the tetrahedral and octahedral holes (see the coordination numbers in Table~\ref{t1}). 
The equilibrium lattice parameters are $a_0^{\rm exp} = 12.232$~{\AA}, 
$b_0^{\rm exp} = 3.041$~{\AA} and 
$c_0^{\rm exp} = 5.801$~{\AA}, 
with $\beta$ = 103.7$^\circ$\cite{geller-676}.
Both types of Ga ions have a formal oxidation state of III, and their lattice
sites contain small lattice distortions away from the ideal polyhedra. A recent hybrid-DFT study has
predicted that formation of an oxygen vacancy in $\beta$-Ga$_2$O$_3$ results
in a deep donor centre\cite{varley-142106}.

\subsection{Al$_2$O$_3$}

$\alpha$-Al$_2$O$_3$ is an important dielectric material\cite{evans-1995}, with a
band gap of 9.25 eV\cite{tomiki-573}, which adopts the hexagonal corundum
mineral structure (space group $R3/c$) with $a_0^{\rm exp} = 4.759$~{\AA} and
$c_0^{\rm exp} = 12.990$~{\AA}\cite{thompson-79}, and octahedral cation coordination. 
It is widely used as an insulating substrate for thin-film growth.
Of particular interest theoretically
has been its defect chemistry\cite{weber-1756, catlow-1006, hine-114111}.
Recent work has focused on the magnetic and spin configuration associated with the oxygen
interstitial\cite{sokol-44}.

\begin{table}[ht]
\caption{\label{t1} Experimental and calculated material properties for a number of binary metal oxides, including the
electronic band gap,
the standard formation enthalpy\cite{crc}, the cation effective coordination number,
and
the Bader partial charge (in electrons).
 }
\begin{ruledtabular}
\begin{tabular}{lccccc}
Compound    					 &      E$_g$ (eV)        				 &    $\Delta H_f$ (eV) & Coordination   &  Partial Charge   \\
\hline
ZnO           				&       3.44\cite{madelung-04}      &     -3.63           & 4.0            &   1.32            \\ 
In$_2$O$_3$   				&       2.90\cite{walsh-167402}     &     -9.60           & 5.9 - 6.0      &   2.02 - 2.03    \\ 
Ga$_2$O$_3$   				&       4.90\cite{orita-4166}				&     -11.29          & 4.0 - 5.8      &   1.80 - 1.91     \\ 
Al$_2$O$_3$   	      &       9.25\cite{tomiki-573}				&     -17.37         	& 6.0            &   2.62             \\      
\end{tabular}
\end{ruledtabular}
\end{table}

\section{Rules governing structure formation}

\subsection{Crystalline multi-component oxides}

\begin{figure}[ht]
\scalebox{0.7}{\includegraphics{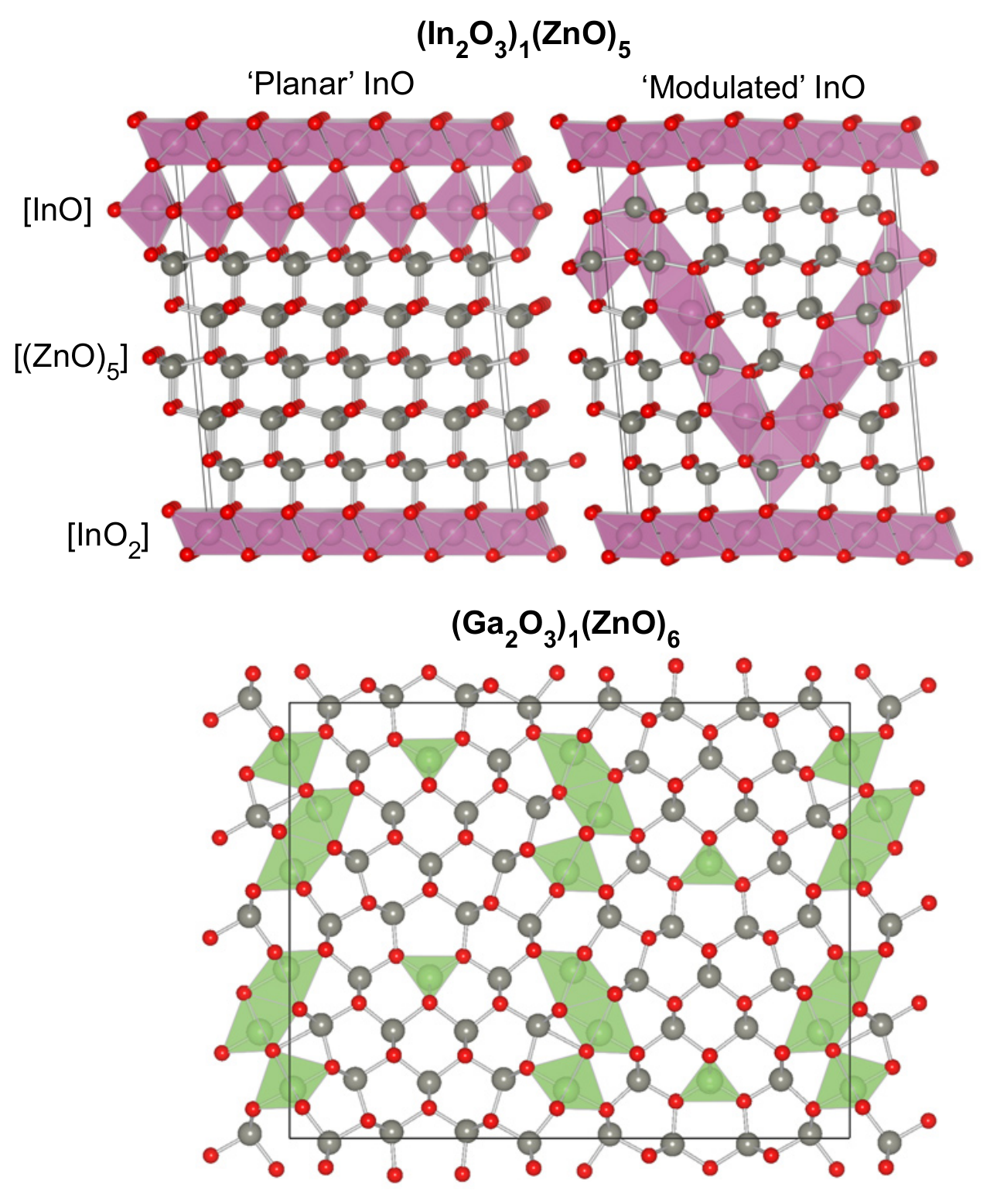}}
\caption{\label{structures} Illustrations of the crystal structures for the planar and modulated models of (In$_2$O$_3$)(ZnO)$_5$ and the lowest energy configuration of (Ga$_2$O$_3$)(ZnO)$_6$. Indium polyhedra are shaded pink, and Ga polyhedra are shaded green, while oxygen and zinc atoms are coloured red and grey, respectively. Visualisation was performed with the VESTA package\cite{vesta}.}
\end{figure}

The multi-compounds ($RM$O$_3$)$_m$(ZnO)$_n$ with $R, M$ = In, Ga, and Al ($n,
m$ = integers) can be synthesized, for example, by solid state chemical reactions using
stoichiometric proportions of the ZnO, In$_2$O$_3$, Ga$_2$O$_3$, Al$_2$O$_3$
binary compounds and heating at high temperatures (about $1000 - 2000$~K). The exact
temperature required depends on the composition and the desired products\cite{Kimizuka-1995-170,Phani-1998-3969,Li-1999-355,Kim-2004-163,Michiue-2008-521}.
A large number of syntheses have been performed; and crystal
structures were first reported for $R$ = In and Ga, with different $M$.
The lowest energy structures identified for a number of these compounds
are shown in Figure \ref{structures}.

%
Except for the spinel structured compound ZnAl$_2$O$_4$, there have been limited reports for the synthesis of single crystal (Al$_2$O$_3$)(ZnO)$_n$ compounds for $n > 1$. The Zn-rich alloy was recently grown by pulsed laser deposition in the work of Yoshioka \textit{et al.}\cite{yoshioka-014309}, where a superlattice structure was identified, which is consistent with the homologous crystal structure models developed for the In and Ga compounds, as later verified by electronic structure calculations\cite{yoshioka-137}. There is growing evidence for the presence of the homologous phase in typical Al-doped ZnO samples\cite{vinnichenko-141907, horwat-132003}, which can contribute to lower than expected electron carrier concentrations: excess electrons are compensated by additional oxygen incorporation in to the lattice.

\subsubsection{(In$_2$O$_3$)$_m$(ZnO)$_n$ and (Ga$_2$O$_3$)$_m$(ZnO)$_n$}

IZO was first synthesized by Kasper\cite{Kasper-1967-113} for $m = 1, n = 2-7$,
who showed that the compounds adopt layered structures and crystallise in
rhombohedral or hexagonal crystal lattices. Two decades later, Cannard and
Tilley confirmed Kasper's results using X-ray diffraction (XRD) and
high-resolution electron microscopy (HREM) experiments\cite{Cannard-1988-418}.
Recently, Kimizuka {\it et al.}\cite{Kimizuka-1981-109,Kimizuka-1995-170}
proposed that IZO is isostructural with LuFeO$_3$(ZnO)$_n$ for $m = 1$
\cite{Isobe-1994-332}. The crystal has the space group $R\bar{3}m$ (rhombohedral
lattice) when $n$ is odd and $P6_3/mmc$ (hexagonal lattice) when $n$ is even.
The In atoms form an InO$_2$ octahedron layer, which are interconnected by
(InZn$_n$)O$_{n+1}$ layers.
The In atoms in the (InZn$_n$)O$_{n+1}$ unit were assumed to be
randomly distributed on the metal sites, forming 5-fold trigonal bipyramidal
polyhedra\cite{Isobe-1994-332}, while the Zn atoms were assumed to be located
in tetrahedral sites surrounded by four O atoms. The formation of an InO$_2$
octahedron layer was also confirmed by atomic-resolution Z-contrast
experiments for In$_2$O$_3$(ZnO)$_n$\cite{Yan-1998-2585}. In contrast to the
XRD studies,\cite{Kimizuka-1981-109,Kimizuka-1995-170,Moriga-1998-1310} recent
HREM experiments\cite{Uchida-1994-146,Li-1998-347,Li-2000-543,Wolf-2007-549}
have indicated that the In atoms in the (InZn$_n$)O$_{n+1}$ layers form
ordered modulated structures with a zig-zag shape.

So far, the synthesis of GZO has been reported only for $m = n = 1$,
\cite{Phani-1998-3969,Kim-2004-163} $m = 1, n = 6$,\cite{Michiue-2008-521} $m
= 1, n = 7, 8, 9, 16$,\cite{Kimizuka-1995-170} and $m = 1, n = 9, 13$.
\cite{Li-1999-355} To our knowledge, there is no report for $m > 1$.
Experimental characterisation using XRD has found that GZO crystallizes in
orthorhombic structures with space group $Cmcm$ for $n > 1$,
 and the  `normal' cubic spinel structure for $m = n = 1$. The internal structural
parameters (atomic positions) have been reported only for $m = 1, n = 1, 6$
\cite{Michiue-2008-521}; however, it is important to note that the atomic
number of Ga and Zn differs only by one unit, and hence, XRD alone cannot clearly
distinguish between the Ga and Zn site occupation, \textit{i.e.} it was assumed that Ga atoms occupy
specific cation sites, while Zn occupy the remaining sites. The pioneering
work reported by Michiue {\it et al.}\cite{Michiue-2008-521} provided a basis
for the understanding of the atomic structure of GZO compounds; however, a number of
questions remained to be solved at that time.

Recently, Da Silva {\it et al.}\cite{dasilva-255501,DaSilva-2009-214118}
employed first-principles computational tools to address the
mechanisms that lead to the formation of the layered IZO  structures with
the In-modulations in the (InZn$_n$)O$_{n+1}$ layers and the atomic distribution of
the Ga and Zn cations in the GZO lattice, which forms natural grain-boundaries.
Below, we will summarize the
most important results, which can be used as guidelines for a
better understanding of these materials and other related oxide compounds.

\paragraph{The preference of In to form octahedron networks:}

The In and O atoms in the bixbyite In$_2$O$_3$ structure are 6- and 4-fold
coordinated,\cite{Marezio-1966-723} respectively, whereas both Zn and O atoms
are 4-fold coordinated in the wz structure.
In contrast to the In atoms, the
Ga atoms in the monoclinic  $\beta$-Ga$_2$O$_3$ structure
are separated into two
groups composed of four  distorted octahedron and four tetrahedron,
while there are eight 3-fold and four 4-fold O atoms. In particular, a
pair of octahedrons are surrounded by 6 tetrahedrons, which can be seen as a
basic motif in $\beta$-Ga$_2$O$_3$. For the simplest multi-compound structure,
\textit{i.e.} the high symmetry $R_2$ZnO$_4$ ($R$ = In, Ga) spinel structure 
may also exist, in which the In, Zn, and O atoms are 6-, 4-, and 4-fold
coordinated, respectively. The coordination environments observed for the In,
Ga, Zn, and O atoms are related to the fact that In, Ga, and Zn have three,
three, and two valence electrons, respectively, whereas oxygen always has a
valence of $-2$, so the coordination environments are determined such that the
octet rule for local charge neutrality is satisfied.
Thus, it can be expected that In should prefer to form
octahedrons, whereas Ga can form both octahedrons as well as
tetrahedrons, even though both atoms have the same formal valence.
To test this hypothesis, we investigate several IZO and GZO model structures.

Two types of 6-fold coordinated InO$_2$ layers were calculated in
IZO, namely, a non-octahedron structure and an octahedron structure. The
octahedron structure is about 1~eV/InO$_2$ lower in energy than the
non-octahedron structure. This is consistent with the fact that the octahedron
structure maximizes the atomic separation between the negatively charged O atoms.
In the octehedron structure, the
In$-$O distances in the  InO$_2$ layers are in the range of $2.20-2.30$~{\AA},
while the angle of O$-$In$-$O deviates by $1-5^{\circ}$ from the ideal value of
$180^{\circ}$. The smallest deviations ($1^{\circ}$) occur when O atoms in the
corner of the octahedron are shared with Zn atoms. Thus, the binding of Zn to
O at the corner of the InO$_2$ octahedron layers stabilises the
formation of an almost perfect In-O octahedron, whereas binding to In
induces a larger distortion. These findings are consistent with the
formation of distorted and perfect octahedron structures in the bixbyite
In$_2$O$_3$ and the spinel In$_2$ZnO$_4$ structures, respectively. The
observation that the In and O atoms in the InO$_2$ layer form an octahedron
structure is consistent with experimental observations.\cite{Kimizuka-1981-109,Giaquinta-1994-5,Isobe-1994-332,Kimizuka-1995-170,Yan-1998-2585}

In contrast to the In atoms in IZO which is 6-fold coordinated, Ga atoms may 
exist at low coordination sites. For example, in the GZO structure proposed 
by Michiue {\it et al.}\cite{Michiue-2008-521} there are two cation sites (called $M8$ and $M7$ in
Ref.~\onlinecite{Michiue-2008-521}), which are 5- and 4-fold coordinated and
were assumed to partially occupied by Zn atoms. We showed that the $M8$ sites
are empty and the $M7$ sites are occupied by Ga atoms, while the remaining Ga
atoms are distributed in 6-fold like sites. Similar to the $\beta$-Ga$_2$O$_3$
structure, both Ga sites satisfy the octet rule. Thus, these results
show clearly that In atoms have a stronger preference for octahedron sites in
IZO than Ga atoms in GZO, which explains the formation of planar InO$_2$
layers in IZO, whereas it is absent in GZO.

\paragraph{Inversion domain boundary formation in ($R$Zn$_n$)O$_{n+1}$
layer:}

As discussed above, the In atoms have a strong preference for
octahedron sites, while the Ga atoms can occupy 6- and 4-fold sites. Thus,
once the 6-fold In sites in the InO$_2$ layer in IZO and the 4-fold Ga sites
in GZO are occupied, the remaining In and Ga atoms must be distributed in the
wz-like ZnO structure; however, all cation sites in the wurtzite structure are
4-fold and not 6-fold. Thus, changes must occur in the ($R$Zn$_n$)O$_{n+1}$
layers in order to accommodate both In and Ga atoms and satisfy the electronic
octet rule.

Furthermore, a number of structural features of IZO and GZO also contribute to
the formation of an inversion domain boundary (IDB). For example, in IZO, the
InO$_2$ octahedron layers are connected by the wz-like
(InZn$_n$)O$_{n+1}$ layers, in which the O atoms at the corner of the
octahedrons are connected to three In atoms in InO$_2$ and one atom in the
(InZn$_n$)O$_{n+1}$ layer. Thus, the polarity at the bottom and top of the
repeating unit is reversed, and hence, an IDB must exist in the
(InZn$_n$)O$_{n+1}$ layer to reverse this polarity. An InO$_2$-like layer is
not present in GZO; however, the distribution of the Ga atoms requires 6-fold
like sites, which can be obtained through the formation of IDB in GZO. For both IZO
and GZO cases, the IDB is located on the In and Ga atoms, which form 5-fold
trigonal bipyramid structures with the surrounding O atoms in the
($R$Zn$_n$)O$_{n+1}$ layers.

\paragraph{Stacking fault in the ($R$Zn$_n$)O$_{n+1}$ layer:}

It is important to realize that the formation of the IDB does not occur alone,
\textit{i.e.} 5-fold trigonal bipyramid sites cannot be created without disrupting the
stacking sequence of the wz-like lattice. In order to preserve the
hexagonality of wurtzite and without
destroying the stable InO$_2$ octahedron layer and the formation of the 4-fold
Ga sites, one or more stacking faults must exist in the ($R$Zn$_n$)O$_{n+1}$
layers. This has important consequences for determining the stable configurations of IZO. 
For example, the
conventional hexagonal unit cells are composed of two, or three, InO$_2$
layers separated by an equal number of (InZn$_n$)O$_{n+1}$ layers in which $n$
is even or odd. This leads to a hexagonal primitive unit cell for $n$ even and
rhombohedral (or monoclinic) for $n$ odd.

\paragraph{Modulation of $R$ atoms in the ($R$Zn$_n$)O$_{n+1}$ layer:}

As previously mentioned, to obey the material stoichiometry, a number of In and Ga atoms
must adopt trigonal bipyramid structures with the surrounding O atoms; at low temperature,
these motifs should form ordered configurations.
The ionic radii of the In, Ga, and Zn atoms,
\textit{i.e.} 0.80~{\AA} for In(III), 0.55~{\AA} for Ga(III), and 0.60~{\AA}
for Zn(II),\cite{shannon-751} play a decisive role in determining the ordered structure
that minimizes the internal strain of the IZO and GZO structures.

For example, model calculations demonstrated that the in-plane lattice
constant of In$M$O$_3$ in the hexagonal InFeO$_3$ structure
\cite{Giaquinta-1994-5} is 8.4\% larger than the in-plane lattice constant of
ZnO in the wz structure. Thus, the In atoms must assume an ordered
distribution in the (InZn$_n$)O$_{n+1}$ layers to minimize the strain energy. 
Calculations for several
different atomic configurations have identified that the In atoms are
distributed in the cation sites following a zig-zag modulated structure in
IZO, in which the modulation period is proportional to the number of ZnO
units. Similarly, the Ga atoms are distributed in GZO to form a modulation, which
match the one observed in IZO.

The modulation is composed of In/Ga and Zn
atoms, which form trigonal bipyramidal structures. These results are
consistent with experimental HREM studies,
\cite{Uchida-1994-146,Li-1998-347,Li-2000-543,Wolf-2007-549} which observed
clearly the formation of In-modulated structures in the (InZn$_n$)$_{n+1}$
layers for $n > 5$ ZnO units and $m = 1$\cite{Li-1998-347}.

\paragraph{Electronic octet rule for the O in the ($R$Zn$_n$)O$_{n+1}$
layers:}

The minimum energy structure for $R$Zn$_n$O$_{n+1}$ obeys the
electronic octet rule (local charge neutrality), in which the presence of
5-fold Zn sites is fundamental for both IZO and GZO. The following
environments were identified in IZO as fundamental for the electronic octet
rule: the O atoms are surrounded by $4\times$Zn(4) atoms, or $2\times$Zn(4) +
$1\times$In(5) + $1\times$Zn(5), or $2\times$In(5) + $2\times$Zn(5), where the
numbers in parentheses indicate the integral coordination of the In and Zn
atoms, so that each O atom always receives two electrons from neighbouring
cations. Similar coordination environments are observed in GZO beyond of the
4-fold Ga sites. Furthermore, the octet rule is also satisfied for oxygen
pairs along the in-plane In and Zn rows,\textit{ e.g.} $2\times{\rm Zn}(4) +
3\times{\rm Zn}(5) + 3\times{\rm In}(5)$, and for O atoms in the InO$_2$
layers. Therefore, the local electrostatic environment plays a
fundamental role in the stability of the ($R_2$O$_3$)$_m$(ZnO)$_n$ compounds.

\subsubsection{(In$M^{III}$O$_3$)$_m$(ZnO)$_n$}

One important requirement for technological applications of IZO is
complete control of the electronic properties, \textit{e.g.} band gap, and hence,
several studies have been performed to obtain a procedure to tune the band gap
of IZO for specific applications. One possible approach is the
replacement of a percentage of In atoms by other trivalent species, \textit{e.g.} Ga or Al atoms.
Therefore, it is very important to understand the consequences of 
such substitutions on the IZO
atomic structure. We observed that the ionic size of the In(III) and
Ga(III) cations relative to Zn(II) plays an important role in the
formation of the $M$-modulations.

In our calculations, we assumed that half
of the In atoms were replaced by Ga or Al atoms to form
(InGaO$_3$)$_m$(ZnO)$_n$ and (InAlO$_3$)$_m$(ZnO)$_n$, respectively.
We found that the In atoms always form the InO$_2$ layer,
which is expected based on the high stability of In octahedron
structures. Replacement of the In atoms in the InO$_2$ layer by Ga or Al atoms
and put In atoms in the ZnO layer, is highly energetically unfavourable.
Thus, Ga and Al should replace the modulated In atoms in the ZnO layers.
We also demonstrated
that the in-plane lattice constant of In$M$O$_3$ in the hexagonal
InFeO$_3$ structure\cite{Giaquinta-1994-5} is larger by 8.4\%, 2.4\%, and
0.0\% for $M$ = In, Ga, and Al, respectively, compared with the lattice
constant of wz-ZnO. Therefore, for $M$ =
In, a large strain exists in the ZnO and InO layers if the In atoms form a
planar layer. The in-plane strain decreases for $M$ = Ga and is almost zero
for $M$ = Al. Therefore, it is probable that the Al atoms are randomly
distributed in IAZO, while in IGZO, modulated Ga ions might still be present.

\subsection{Amorphous multi-component oxides}

In contrast to the high temperature associated with the synthesis of crystalline multi-component oxides, amorphous thin films can be grown by low temperature processing. Following an initial investigation by Hosono \textit{et al.} on amorphous ternary oxides\cite{hosono-165}, the breakthrough for a-TCOs came from the report by Nomura \textit{et al.}\cite{nomura-488} that the electron mobility of amorphous IGZO thin films could exceed 10 cm$^2$V$^{-1}$s$^{-1}$, which is an order of magnitude greater than amorphous Si. IGZO has since become the most widely studied amorphous TCO, and has found application as an electron-injection-layer for solid-state lighting devices\cite{nomura-488, nomura-1269, nomura-202117}.

\begin{figure}[ht]
\scalebox{0.47}{\includegraphics{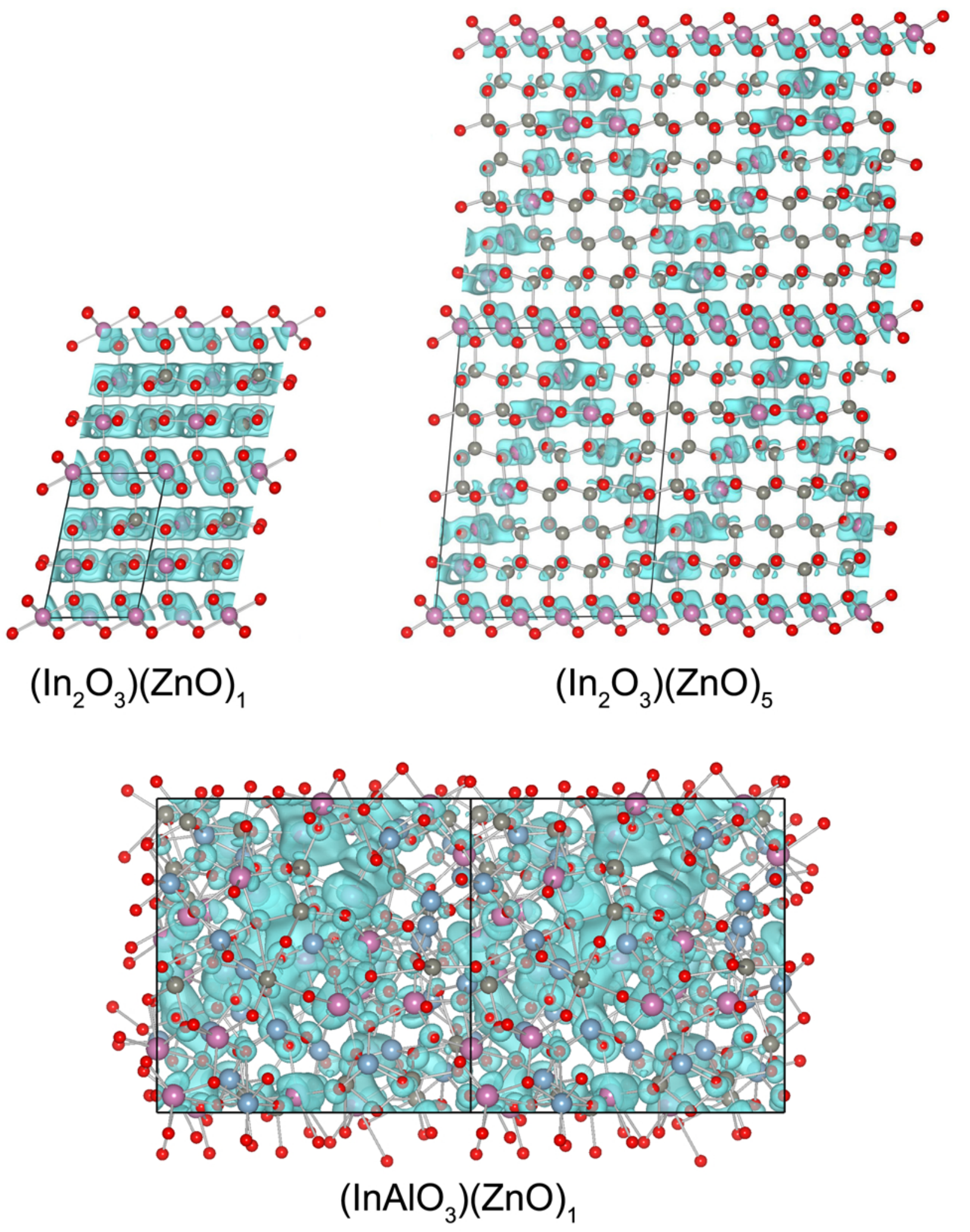}}
\caption{\label{electrons} Conduction band charge density isosurfaces (coloured turquoise) for two crystalline and one amorphous oxide material. The oxygen ions are coloured red.}
\end{figure}

To simulate these low symmetry structures, a pseudo-cubic supercell of the crystalline structures was generated, which was subject to high temperatures under  {\it ab initio} molecular dynamics until a solid to liquid phase transition was observed. Then, the liquid oxide was quenched, and the resulting structures were subject to local optimisation of both the internal positions and lattice constant. The full details of the approach employed can be found in Reference \onlinecite{walsh-5119}. A snapshot of an amorphous structure is shown in Figure \ref{electrons}.

As previously shown, the crystal structures of the In and Zn containing oxides are defined by mixed In-O and Zn-O polyhedra (bipyramids and tetrahedra) present between InO$_2$ octahedral planes. In the amorphous phases of IZO, IGZO and IAZO, the same local coordination motifs are maintained. To compare the local order in both phases, we have calculated both the effective coordination number\cite{ecn-1,ecn-2} and effective charge distribution for each ion\cite{walsh-5119}.
Even in the amorphous phase, the strong charge transfer between the metal cations and oxygen is preserved, driving the similar local coordination. While this is to be expected due to the large electro-negativity of oxygen, it is in contrast to covalent semiconductors, where amorphisation can lead to a large and significant local bond rearrangements that perturb the nearest neighbour environments\cite{wooten-1392}.

The total energy difference between the crystalline and amorphous TCOs is of the order of 200 meV per f.u. in each system studied, which is expected based on the increased strain due to the disordered packing of the cation centred polyhedra on amorphisation. Despite their high energy, the a-TCOs remain metastable experimentally at standard temperatures as they do not have sufficient energy to overcome the large kinetic barriers required for recrystallization.

\section{Optical and Electronic Properties}

The performance of electroceramics for TCO applications can be quantified by their electrical conductivity and optical absorption spectra. The former is related phenomenologically to the position of the electronic states on an absolute energy scale (band alignment), which influences the material's doping limits, while the latter depends on the magnitude of the fundamental band gap, dipole selection rules for optical transitions, and carrier concentrations (\textit{e.g.} the Moss-Burstein effect\cite{moss-775, burstein-632}).

Beyond the bulk material properties, the role of defects and non-stoichiometry in these oxides is also important in determining their performance for TCO applications. The theoretical framework for the accurate modelling of point defect formation exists, and the resulting free energies of formation for each defect charge state can be solved self-consistently to produce temperature dependant equilibrium defect and carrier concentrations\cite{kroger-1974, smith-2000}. Quantitative defect modelling techniques have not yet been applied to the case of the complex multi-component systems. The difficulty lies in the large number of inequivalent lattice positions, which must be calculated independently, as well as the lower lattice symmetry in comparison to the binary components; however, with the increase of high performance computing resources to the peta-scale, these issues will soon be overcome. Here, we can rationalise the material performance based on the bulk electronic changes observed in the multi-component systems.

\subsection{Natural band offsets}
The band offsets between semiconductors is one of the most fundamental properties for device physics\cite{yu-05}. The absolute band edge positions of a material have a role in determining intrinsic doping limits,\cite{walukiewicz-123,zhang-1232} as well as their electrochemical properties\cite{xu-543} and interfacial electron transport. Experimentally, X-ray Photoemission Spectroscopy (XPS) can be used to measure the position of atomic-like core states, which provide a reference for the alignment between the valence bands of two or more systems\cite{kowalczyk-1605}. A theoretical approach has been developed in the same spirit, from which the `natural' valence band offset of two compounds at their respective equilibrium positions can be calculated\cite{wei-2011, wei-144}; this approach has recently been revised to take into account the deformation of the core states with respect to volume changes\cite{li-212109}. The conduction band offsets can be inferred by adding the fundamental band gap to the computed valence band offset.

\begin{figure*}[ht]
\scalebox{1.0}{\includegraphics{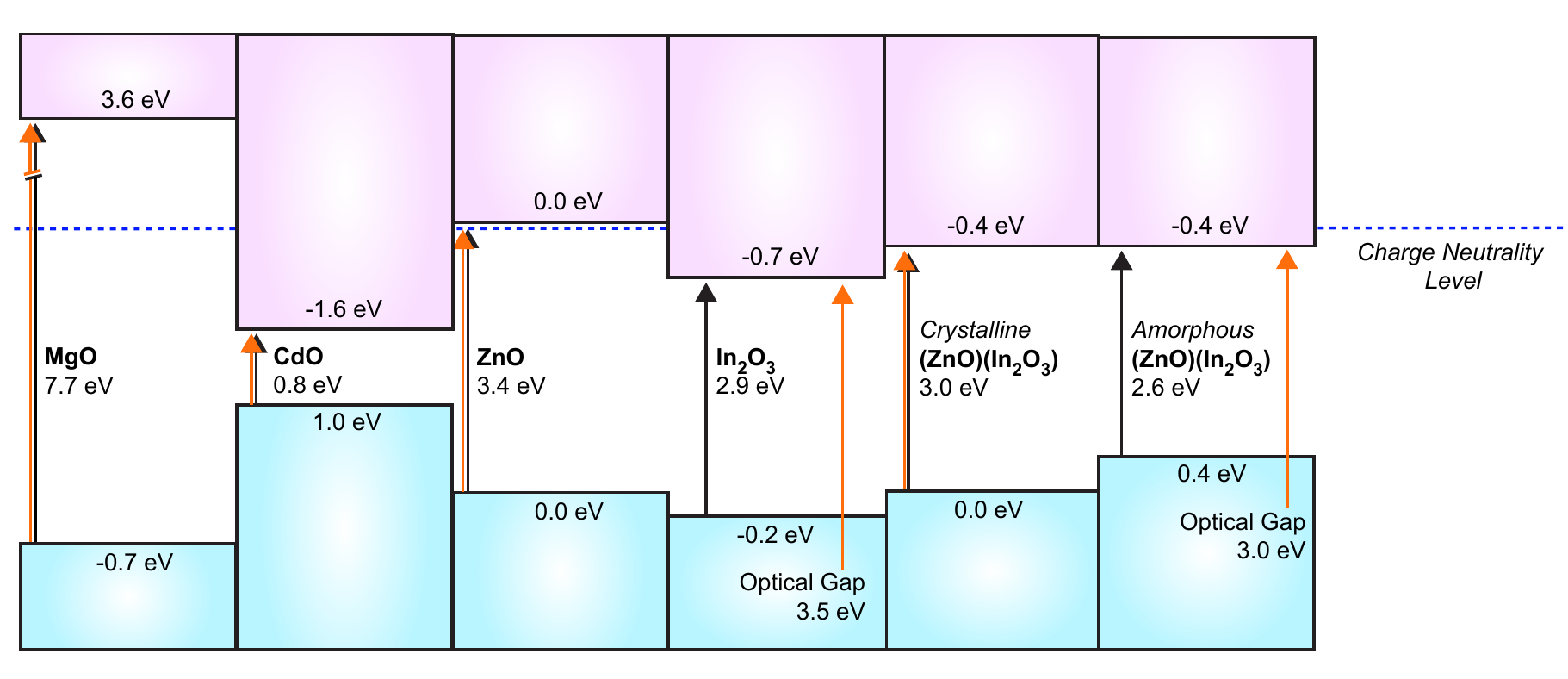}}
\caption{\label{offset} Natural band alignment between In$_2$O$_3$, ZnO and related oxides, relative to that of wurtzite ZnO, and compiled from References \onlinecite{zhu-245209, walsh-5119, walsh-073105, li-212109}. The differences between the electronic and optical band gaps of In$_2$O$_3$ and amorphous (ZnO)(In$_2$O$_3$) are emphasised. The placement of the charge neutrality level is based on an explicit calculation for In$_2$O$_3$.}
\end{figure*}

The natural band offsets derived for a number of oxide materials of interest are shown in Figure \ref{offset}. The results match the recent alignment of the ultraviolet and inverse photoemission spectra for ZnO and In$_2$O$_3$\cite{kamiya-1061}. The observed trends are determined by a combination of shallow cation $d$ states (\textit{e.g.} Cd 4$d$ states raise the valence band of CdO), the degree of ionicity (\textit{e.g.} efficient electron transfer produces a high electrostatic potential that lowers the valence band of MgO), and crystal structure (\textit{e.g.} low coordination oxygen sites raise the valence band for amorphous IZO).

It can be observed from  Figure \ref{offset} that In$_2$O$_3$ has the lowest conduction band level (highest electron affinity) of all the transparent oxides studied, which correlates with its ability to sustain high electron carrier concentrations that result in degenerate electron conduction. Indeed, recent experimental work has shown that the charge neutrality level, \textit{i.e.} the Fermi energy at which oppositely charged defects can form spontaneously, of In$_2$O$_3$ lies well above the conduction band\cite{king-116808}, which has been supported by calculations at the level of many-body perturbation theory\cite{schleife-012104}; further detail and implications of the charge neutrality level are discussed in work by King in this themed issue\cite{king-jpcm}. While ZnO is a good $n$-type semiconductor, the higher conduction band level limits its ability to be heavily electron doped. For large gap oxides such as Al$_2$O$_3$ and Ga$_2$O$_3$, electron doping will be even more difficult, with high degrees of ionic charge compensation expected to occur\cite{catlow-nm}. These effects have been summarised in a series of doping limit rules for semiconductors\cite{zhang-1232, wei-337, lany-045501}.

\subsection{Forbidden optical transitions and carrier localisation in (In$_2$O$_3$)(ZnO)$_n$}

The principal reason why In$_2$O$_3$ remains a good TCO, despite its relatively small band gap, is that the band edge optical transitions are symmetry forbidden: it combines a small fundamental gap with a large optical band gap. Can this also be the case for the ternary In-containing compounds? For the homologous (In$_2$O$_3$)(ZnO)$_n$ series, we have shown that the lowest energy optical transitions from the valence to conduction bands are dipole allowed for all values of $n$, as the top of the valence band state has large contribution from ZnO \cite{walsh-073105}. This explains the anomaly of why the apparent optical gap of In$_2$O$_3$ (\textit{ca.} 3.6 eV)\cite{hamberg-3240} reduces to below 3.2 eV on the addition of ZnO to form IZO\cite{minami-971, moriga-312, taylor-90}.

For the IZO compounds, while the top of the valence band is more localised on the Zn-O network, the conduction band is localised on the In-O network, which follows from the band alignment of the binary materials (Figure \ref{offset}). For $n$ = 1, the localisation is not significant as there are two mixed cation layers present between the In octahedra, and these contain an equal number of In and Zn ions. For higher values of $n$, the number of intermixed layers increases, which raises the Zn:In ratio, and the band edge wavefunctions become more localised as a result. With the increased Zn concentration, an effective superlattice is formed and the charge carriers become confined in the InO$_2$ octahedron network. The modulation of In ions through the ZnO layers makes a significant additional contribution to the conduction band, producing a modulated electron density distribution, as illustrated in Figure \ref{electrons}. The predicted carrier localization correlates well with the minimum of resistivity previously found for IZO samples with low Zn:In ratio, \textit{i.e.} for the range $0 < n < 1$\cite{minami-971, moriga-312, taylor-90}.

\subsection{High performance of amorphous oxides}

For the crystalline TCO compounds, the band gaps follow the clear trend of the group 13 binary oxides, increasing from In to Ga to Al\cite{walsh-5119}. For the amorphous compounds, the same band gap trend is observed, but the magnitudes are lower compared to the crystalline structures. The differences arise from the series of bands situated at the top of the valence band. In the crystalline oxides, the lowest binding energy valence bands are dominated by O 2$p$ states, and this is unchanged for the amorphous systems; however, the variation in electrostatic potential around each of the oxygen sites in the amorphous structures results in the spatial localisation of the valence wavefunctions. The contribution of the O 2$p$ valence band to the sub band gap states has been confirmed through hard X-ray photoemission measurements\cite{nomura-202117}, and is consistent with the band gap decrease observed for a-TCOs\cite{kumar-073703}. As these states are concentrated in the  Zn-rich regions (due to the lower binding energy shallow-core Zn 3$d$  states), their contribution to the absorption of visible light is not significant because the conduction band is more localised on the group 13 cation side, and the resulting optical transition probabilities are low. As a result, the amorphous materials retain sufficient transparency in the visible region to be viable for TCO applications.

The electronic density of states predicted for the amorphous oxides has important consequences. Crucially for $n$-type conductivity, the conduction band remains highly delocalised, which can be understood from the overlap between the disperse cation $s$ orbitals and the isotropic nature of the cation $s$ $-$ O 2$s$ bonding\cite{nomura-488, medvedeva-57004, robertson-1026}: the overlap is largely unaffected by the long range structural disorder, which is illustrated in Figure \ref{electrons}.

 As expected from their larger In$-$O bond length compared to the other cations, In makes the largest contribution to the antibonding conduction band state. For IZO, two thirds of the cations are In, while for IGZO and IAZO, only one third In are present. As stated earlier, even for the lower concentration materials such as IAZO, the In$-$O polyhedra form a combination of edge and corner sharing networks ensuring that the conduction band is appropriately spread over the entire amorphous network. This also explains why the amorphous, low indium concentration TCOs can exhibit good electrical transport properties: in the crystalline compounds, the In$-$O networks become confined to two dimensional planes separated by the Zn and group 13 cations, while in the amorphous compounds, the indium ions become evenly distributed and hence offer improved isotropic carrier mobility.

While a-TCOs exhibit excellent $n$-type conductivity, $p$-type behaviour is not expected to be present due to the strong localisation of the upper valence band. The majority of known $p$-type oxide semiconductors such CuAlO$_2$ and SrCu$_2$O$_2$ are derived from Cu$_2$O, where electron deficiency (hole formation) is facilitated by Cu oxidation, with carrier mobility along the linear O-Cu-O chains. Here, electrical carrier (polaron) mobility will be inhibited by the structural disorder on amorphisation. The absence of $p$-type response has been reported experimentally\cite{nomura-202117}. The discovery of a high performance $p$-type TCO material, either amorphous or crystalline, remains an open challenge in this field.

\section{Discussion and Outlook}

It is clear that the formation of multi-component oxides offers a way to tune the physicochemical properties associated with transparency and conduction.
We have highlighted a number of rules that determine the structure formation of these complex materials.
While the crystal structures of the binary oxide components are well described by the established principles of Pauling\cite{pauling-1010, pauling-1960}, for the multi-cation systems,
features such as polarity inversion, stacking faults and partial site occupation arise.
In terms of the electronic properties, for these $n$-type materials, the key attribute is a delocalised conduction state,
which can be maintained when a significant amount of In is present.
The mixing of Ga and Al serves to both increase the band gap and the crystal stability,
which can provide a means to control the stoichiometry and optical transmission, \textit{e.g.}
to provide a blue-shift in the optical band gap suitable for UV applications.

Despite these advances, a number of issues are still outstanding in the field of multi-component TCOs.
Even in the `simpler' binary metal oxide components, the origin and nature of conductivity remains a contentious issue. One common feature is that oxygen vacancies appear to be deep donor centres, with the neutral defects being stabilized by the strong Madelung potential of the lattice; conversely, metal interstitials generally act as shallow donors, but have higher formation energies. The role of hydrogen as an extrinsic donor impurity has been proposed for a number of TCO materials\cite{cox-2601, janotti-165202, king-081201, king-062110}, but it is highly unlikely to be the primary electron source. Fortunately, atomistic simulation techniques are becoming increasingly quantitative, and we can now predict defect formation energies with high accuracy, especially when temperature dependent contributions to the free energy are included. It is therefore anticipated that our understanding of the binary oxide systems will be solved in the near future, and that this understanding will transfer to the more complex multi-component systems.

One real limitation to transparent electronics is the absence of $p$-type TCO materials, which are needed to form $p-n$ junctions for diodes and transistors. It appears that the performance of Cu-based compounds will not be sufficient for commercial applications. An alternative route to the use of Cu(I) is to form oxides of cations with filled $s$ bands such as  Sn(II) or Bi(III)\cite{walsh-547}. Both SnO and BiVO$_4$ have attracted recent attention for their $p$-type conductivity\cite{ogo-032113,vinke-83}. Moving away from pure oxide materials, forming solid-solutions with chalcogenides (S, Se and Te) may offer a way to increase hole stability by raising the valence band energy. In addition, the formation of hybrid materials containing $n$-type inorganic networks and $p$-type organic networks is also feasible\cite{cheetham-58}, and we have performed exploratory work in this area\cite{walsh-1284, walsh-2341}.

\section{Conclusion}
We have demonstrated the success of computational modelling techniques applied to the complex case of multi-component transparent conducting oxides. In the last decade, our understanding of the structural, electronic and optical properties of these materials has increased significantly. A number of fundamental `rules' have been established relating to crystal structure formation, as well as the conductivity trends, which can provide a framework to enable, for example, a high-throughput computational screening of candidate compounds and stoichiometries. We note that despite these advances, many open questions remain in the field, especially concerning the microscopic origin of conductivity and the role of extended defects, such as dislocations and grain boundaries, which ensure that the field will remain active for the near future.

~

\begin{acknowledgements}
A.W. would like to acknowledge funding from a Marie-Curie Intra-European Fellowship from the European Union under the Seventh Framework Programme, as well as membership of the UK's HPC Materials Chemistry Consortium, which is funded by EPSRC (Grant No. EP/F067496).
J.L.F.D.S.  thanks the S\~ao Paulo Science Foundation (FAPESP).
S.-H.W. is supported by the U.S. Department of Energy (DOE) under Contract No. DE-AC36-08GO28308; computing resources of the National Energy Research Scientific Computing Center were employed, which is supported by DOE under Contract No. DE-AC02-05CH11231.
\end{acknowledgements}


\begin{thebibliography}{100}%
\makeatletter
\providecommand \@ifxundefined [1]{%
 \ifx #1\undefined \expandafter \@firstoftwo
 \else \expandafter \@secondoftwo
\fi
}%
\providecommand \@ifnum [1]{%
 \ifnum #1\expandafter \@firstoftwo
 \else \expandafter \@secondoftwo
\fi
}%
\providecommand \enquote [1]{``#1''}%
\providecommand \bibnamefont  [1]{#1}%
\providecommand \bibfnamefont [1]{#1}%
\providecommand \citenamefont [1]{#1}%
\providecommand\href[0]{\@sanitize\@href}%
\providecommand\@href[1]{\endgroup\@@startlink{#1}\endgroup\@@href}%
\providecommand\@@href[1]{#1\@@endlink}%
\providecommand \@sanitize [0]{\begingroup\catcode`\&12\catcode`\#12\relax}%
\@ifxundefined \pdfoutput {\@firstoftwo}{%
 \@ifnum{\z@=\pdfoutput}{\@firstoftwo}{\@secondoftwo}%
}{%
 \providecommand\@@startlink[1]{\leavevmode}%
 \providecommand\@@endlink[0]{}%
}{%
 \providecommand\@@startlink[1]{%
  \leavevmode
  \pdfstartlink
   attr{/Border[0 0 1 ]/H/I/C[0 1 1]}%
   user{/Subtype/Link/A<</Type/Action/S/URI/URI(#1)>>}%
  \relax
 }%
 \providecommand\@@endlink[0]{\pdfendlink}%
}%
\providecommand \url  [0]{\begingroup\@sanitize \@url }%
\providecommand \@url [1]{\endgroup\@href {#1}{\urlprefix}}%
\providecommand \urlprefix [0]{URL }%
\providecommand \Eprint[0]{\href }%
\@ifxundefined \urlstyle {%
  \providecommand \doi [1]{doi:\discretionary{}{}{}#1}%
}{%
  \providecommand \doi [0]{doi:\discretionary{}{}{}\begingroup
  \urlstyle{rm}\Url }%
}%
\providecommand \doibase [0]{http://dx.doi.org/}%
\providecommand \Doi[1]{\href{\doibase#1}}%
\providecommand \bibAnnote [3]{%
  \BibitemShut{#1}%
  \begin{quotation}\noindent
    \textsc{Key:}\ #2\\\textsc{Annotation:}\ #3%
  \end{quotation}%
}%
\providecommand \bibAnnoteFile [2]{%
  \IfFileExists{#2}{\bibAnnote {#1} {#2} {\input{#2}}}{}%
}%
\providecommand \typeout [0]{\immediate \write \m@ne }%
\providecommand \selectlanguage [0]{\@gobble}%
\providecommand \bibinfo [0]{\@secondoftwo}%
\providecommand \bibfield [0]{\@secondoftwo}%
\providecommand \translation [1]{[#1]}%
\providecommand \BibitemOpen[0]{}%
\providecommand \bibitemStop [0]{}%
\providecommand \bibitemNoStop [0]{.\EOS\space}%
\providecommand \EOS [0]{\spacefactor3000\relax}%
\providecommand \BibitemShut [1]{\csname bibitem#1\endcsname}%
\bibitem{thomas-907}%
  \BibitemOpen
  \bibfield{author}{%
  \bibinfo {author} {\bibfnamefont{G.}~\bibnamefont{Thomas}},\ }%
  \bibfield{journal}{%
  \Doi{10.1038/39999}{\bibinfo {journal} {Nature}}\ }%
  \textbf{\bibinfo {volume} {389}},\ \bibinfo {pages} {907} (\bibinfo {year}
  {1997})%
  \bibAnnoteFile{NoStop}{thomas-907}%
\bibitem{edwards-2295}%
  \BibitemOpen
  \bibfield{author}{%
  \bibinfo {author} {\bibfnamefont{P.~P.}\ \bibnamefont{Edwards}}, \bibinfo
  {author} {\bibfnamefont{A.}~\bibnamefont{Porch}}, \bibinfo {author}
  {\bibfnamefont{M.~O.}\ \bibnamefont{Jones}}, \bibinfo {author}
  {\bibfnamefont{D.~V.}\ \bibnamefont{Morgan}},\ and\ \bibinfo {author}
  {\bibfnamefont{R.~M.}\ \bibnamefont{Perks}},\ }%
  \bibfield{journal}{%
  \bibinfo {journal} {Dalton Trans.}\ }%
  \textbf{\bibinfo {volume} {15}},\ \bibinfo {pages} {2295} (\bibinfo {year}
  {2004})%
  \bibAnnoteFile{NoStop}{edwards-2295}%
\bibitem{ohya-240}%
  \BibitemOpen
  \bibfield{author}{%
  \bibinfo {author} {\bibfnamefont{Y.}~\bibnamefont{Ohya}}, \bibinfo {author}
  {\bibfnamefont{T.}~\bibnamefont{Yamamoto}},\ and\ \bibinfo {author}
  {\bibfnamefont{T.}~\bibnamefont{Ban}},\ }%
  \bibfield{journal}{%
  \bibinfo {journal} {J. Am. Ceram. Soc.}\ }%
  \textbf{\bibinfo {volume} {91}},\ \bibinfo {pages} {240} (\bibinfo {year}
  {2008})%
  \bibAnnoteFile{NoStop}{ohya-240}%
\bibitem{wit-143}%
  \BibitemOpen
  \bibfield{author}{%
  \bibinfo {author} {\bibfnamefont{J.~H.~W.}\ \bibnamefont{DeWit}},\ }%
  \bibfield{journal}{%
  \bibinfo {journal} {J. Solid State Chem.}\ }%
  \textbf{\bibinfo {volume} {20}},\ \bibinfo {pages} {143} (\bibinfo {year}
  {1977})%
  \bibAnnoteFile{NoStop}{wit-143}%
\bibitem{wit-142}%
  \BibitemOpen
  \bibfield{author}{%
  \bibinfo {author} {\bibfnamefont{J.~H.~W.}\ \bibnamefont{DeWit}},\ }%
  \bibfield{journal}{%
  \bibinfo {journal} {J. Solid State Chem.}\ }%
  \textbf{\bibinfo {volume} {8}},\ \bibinfo {pages} {142} (\bibinfo {year}
  {1973})%
  \bibAnnoteFile{NoStop}{wit-142}%
\bibitem{weiher-2834}%
  \BibitemOpen
  \bibfield{author}{%
  \bibinfo {author} {\bibfnamefont{R.~L.}\ \bibnamefont{Weiher}},\ }%
  \bibfield{journal}{%
  \bibinfo {journal} {J. Appl. Phys.}\ }%
  \textbf{\bibinfo {volume} {33}},\ \bibinfo {pages} {2834} (\bibinfo {year}
  {1962})%
  \bibAnnoteFile{NoStop}{weiher-2834}%
\bibitem{hamberg-3240}%
  \BibitemOpen
  \bibfield{author}{%
  \bibinfo {author} {\bibfnamefont{I.}~\bibnamefont{Hamberg}}, \bibinfo
  {author} {\bibfnamefont{C.~G.}\ \bibnamefont{Granqvist}}, \bibinfo {author}
  {\bibfnamefont{K.~F.}\ \bibnamefont{Berggren}}, \bibinfo {author}
  {\bibfnamefont{B.~E.}\ \bibnamefont{Sernelius}},\ and\ \bibinfo {author}
  {\bibfnamefont{L.}~\bibnamefont{Engstrom}},\ }%
  \bibfield{journal}{%
  \bibinfo {journal} {Phys. Rev. B}\ }%
  \textbf{\bibinfo {volume} {30}},\ \bibinfo {pages} {3240} (\bibinfo {year}
  {1984})%
  \bibAnnoteFile{NoStop}{hamberg-3240}%
\bibitem{king-116808}%
  \BibitemOpen
  \bibfield{author}{%
  \bibinfo {author} {\bibfnamefont{P.~D.~C.}\ \bibnamefont{King}}, \bibinfo
  {author} {\bibfnamefont{T.~D.}\ \bibnamefont{Veal}}, \bibinfo {author}
  {\bibfnamefont{D.~J.}\ \bibnamefont{Payne}}, \bibinfo {author}
  {\bibfnamefont{A.}~\bibnamefont{Bourlange}}, \bibinfo {author}
  {\bibfnamefont{R.~G.}\ \bibnamefont{Egdell}},\ and\ \bibinfo {author}
  {\bibfnamefont{C.~F.}\ \bibnamefont{McConville}},\ }%
  \bibfield{journal}{%
  \Doi{10.1103/PhysRevLett.101.116808}{\bibinfo {journal} {Phys. Rev. Lett.}}\
  }%
  \textbf{\bibinfo {volume} {101}},\ \bibinfo {pages} {116808} (\bibinfo {year}
  {2008})%
  \bibAnnoteFile{NoStop}{king-116808}%
\bibitem{korber-165207}%
  \BibitemOpen
  \bibfield{author}{%
  \bibinfo {author} {\bibfnamefont{C.}~\bibnamefont{K\"orber}}, \bibinfo
  {author} {\bibfnamefont{V.}~\bibnamefont{Krishnakumar}}, \bibinfo {author}
  {\bibfnamefont{A.}~\bibnamefont{Klein}}, \bibinfo {author}
  {\bibfnamefont{G.}~\bibnamefont{Panaccione}}, \bibinfo {author}
  {\bibfnamefont{P.}~\bibnamefont{Torelli}}, \bibinfo {author}
  {\bibfnamefont{A.}~\bibnamefont{Walsh}}, \bibinfo {author}
  {\bibfnamefont{J.~L.~F.}\ \bibnamefont{Da~Silva}}, \bibinfo {author}
  {\bibfnamefont{S.-H.}\ \bibnamefont{Wei}}, \bibinfo {author}
  {\bibfnamefont{R.~G.}\ \bibnamefont{Egdell}},\ and\ \bibinfo {author}
  {\bibfnamefont{D.~J.}\ \bibnamefont{Payne}},\ }%
  \bibfield{journal}{%
  \Doi{10.1103/PhysRevB.81.165207}{\bibinfo {journal} {Phys. Rev. B}}\ }%
  \textbf{\bibinfo {volume} {81}},\ \bibinfo {pages} {165207} (\bibinfo {month}
  {Apr}\ \bibinfo {year} {2010})%
  \bibAnnoteFile{NoStop}{korber-165207}%
\bibitem{klein-1197}%
  \BibitemOpen
  \bibfield{author}{%
  \bibinfo {author} {\bibfnamefont{A.}~\bibnamefont{Klein}}, \bibinfo {author}
  {\bibfnamefont{C.}~\bibnamefont{Korber}}, \bibinfo {author}
  {\bibfnamefont{A.}~\bibnamefont{Wachau}}, \bibinfo {author}
  {\bibfnamefont{F.}~\bibnamefont{Sauberlich}}, \bibinfo {author}
  {\bibfnamefont{Y.}~\bibnamefont{Gassenbauer}}, \bibinfo {author}
  {\bibfnamefont{R.}~\bibnamefont{Schafranek}}, \bibinfo {author}
  {\bibfnamefont{S.}~\bibnamefont{Harvey}},\ and\ \bibinfo {author}
  {\bibnamefont{T.O.Mason}},\ }%
  \bibfield{journal}{%
  \bibinfo {journal} {Thin Solid Films}\ }%
  \textbf{\bibinfo {volume} {518}},\ \bibinfo {pages} {1197} (\bibinfo {year}
  {2009})%
  \bibAnnoteFile{NoStop}{klein-1197}%
\bibitem{walsh-167402}%
  \BibitemOpen
  \bibfield{author}{%
  \bibinfo {author} {\bibfnamefont{A.}~\bibnamefont{Walsh}}, \bibinfo {author}
  {\bibfnamefont{J.~L.~F.}\ \bibnamefont{Da~Silva}}, \bibinfo {author}
  {\bibfnamefont{S.-H.}\ \bibnamefont{Wei}}, \bibinfo {author}
  {\bibfnamefont{C.}~\bibnamefont{K\"orber}}, \bibinfo {author}
  {\bibfnamefont{A.}~\bibnamefont{Klein}}, \bibinfo {author}
  {\bibfnamefont{L.~F.~J.}\ \bibnamefont{Piper}}, \bibinfo {author}
  {\bibfnamefont{A.}~\bibnamefont{DeMasi}}, \bibinfo {author}
  {\bibfnamefont{K.~E.}\ \bibnamefont{Smith}}, \bibinfo {author}
  {\bibfnamefont{G.}~\bibnamefont{Panaccione}}, \bibinfo {author}
  {\bibfnamefont{P.}~\bibnamefont{Torelli}}, \bibinfo {author}
  {\bibfnamefont{D.~J.}\ \bibnamefont{Payne}}, \bibinfo {author}
  {\bibfnamefont{A.}~\bibnamefont{Bourlange}},\ and\ \bibinfo {author}
  {\bibfnamefont{R.~G.}\ \bibnamefont{Egdell}},\ }%
  \bibfield{journal}{%
  \Doi{10.1103/PhysRevLett.100.167402}{\bibinfo {journal} {Phys. Rev. Lett.}}\
  }%
  \textbf{\bibinfo {volume} {100}},\ \bibinfo {pages} {167402} (\bibinfo {year}
  {2008})%
  \bibAnnoteFile{NoStop}{walsh-167402}%
\bibitem{walsh-075211}%
  \BibitemOpen
  \bibfield{author}{%
  \bibinfo {author} {\bibfnamefont{A.}~\bibnamefont{Walsh}}, \bibinfo {author}
  {\bibfnamefont{J.~L.~F.}\ \bibnamefont{Da~Silva}},\ and\ \bibinfo {author}
  {\bibfnamefont{S.-H.}\ \bibnamefont{Wei}},\ }%
  \bibfield{journal}{%
  \Doi{10.1103/PhysRevB.78.075211}{\bibinfo {journal} {Phys. Rev. B}}\ }%
  \textbf{\bibinfo {volume} {78}},\ \bibinfo {pages} {075211} (\bibinfo {year}
  {2008})%
  \bibAnnoteFile{NoStop}{walsh-075211}%
\bibitem{walsh-10438}%
  \BibitemOpen
  \bibfield{author}{%
  \bibinfo {author} {\bibfnamefont{A.}~\bibnamefont{Walsh}}\ and\ \bibinfo
  {author} {\bibfnamefont{C.~R.~A.}\ \bibnamefont{Catlow}},\ }%
  \bibfield{journal}{%
  \bibinfo {journal} {J. Mater. Chem.}\ }%
  \textbf{\bibinfo {volume} {20}},\ \bibinfo {pages} {10438} (\bibinfo {year}
  {2010})%
  \bibAnnoteFile{NoStop}{walsh-10438}%
\bibitem{lany-045501}%
  \BibitemOpen
  \bibfield{author}{%
  \bibinfo {author} {\bibfnamefont{S.}~\bibnamefont{Lany}}\ and\ \bibinfo
  {author} {\bibfnamefont{A.}~\bibnamefont{Zunger}},\ }%
  \bibfield{journal}{%
  \bibinfo {journal} {Phys. Rev. Lett.}\ }%
  \textbf{\bibinfo {volume} {98}},\ \bibinfo {pages} {045501} (\bibinfo {year}
  {2007})%
  \bibAnnoteFile{NoStop}{lany-045501}%
\bibitem{agoston-455801}%
  \BibitemOpen
  \bibfield{author}{%
  \bibinfo {author} {\bibfnamefont{P.}~\bibnamefont{\'Agoston}}, \bibinfo
  {author} {\bibfnamefont{P.}~\bibnamefont{Erhart}}, \bibinfo {author}
  {\bibfnamefont{A.}~\bibnamefont{Klein}},\ and\ \bibinfo {author}
  {\bibfnamefont{K.}~\bibnamefont{Albe}},\ }%
  \bibfield{journal}{%
  \bibinfo {journal} {J. Phys.: Condens. Matter}\ }%
  \textbf{\bibinfo {volume} {21}},\ \bibinfo {pages} {455801} (\bibinfo {year}
  {2009})%
  \bibAnnoteFile{NoStop}{agoston-455801}%
\bibitem{tomita-051911}%
  \BibitemOpen
  \bibfield{author}{%
  \bibinfo {author} {\bibfnamefont{T.}~\bibnamefont{Tomita}}, \bibinfo {author}
  {\bibfnamefont{K.}~\bibnamefont{Yamashita}}, \bibinfo {author}
  {\bibfnamefont{Y.}~\bibnamefont{Hayafuji}},\ and\ \bibinfo {author}
  {\bibfnamefont{H.}~\bibnamefont{Adachi}},\ }%
  \bibfield{journal}{%
  \bibinfo {journal} {Appl. Phys. Lett.}\ }%
  \textbf{\bibinfo {volume} {87}},\ \bibinfo {pages} {051911} (\bibinfo {year}
  {2005})%
  \bibAnnoteFile{NoStop}{tomita-051911}%
\bibitem{agoston-245501}%
  \BibitemOpen
  \bibfield{author}{%
  \bibinfo {author} {\bibfnamefont{P.}~\bibnamefont{\'Agoston}}, \bibinfo
  {author} {\bibfnamefont{K.}~\bibnamefont{Albe}}, \bibinfo {author}
  {\bibfnamefont{R.~M.}\ \bibnamefont{Nieminen}},\ and\ \bibinfo {author}
  {\bibfnamefont{M.~J.}\ \bibnamefont{Puska}},\ }%
  \bibfield{journal}{%
  \bibinfo {journal} {Phys. Rev. Lett.}\ }%
  \textbf{\bibinfo {volume} {103}},\ \bibinfo {pages} {245501} (\bibinfo {year}
  {2009})%
  \bibAnnoteFile{NoStop}{agoston-245501}%
\bibitem{medvedeva-125116}%
  \BibitemOpen
  \bibfield{author}{%
  \bibinfo {author} {\bibfnamefont{J.~E.}\ \bibnamefont{Medvedeva}}\ and\
  \bibinfo {author} {\bibfnamefont{C.~L.}\ \bibnamefont{Hettiarachchi}},\ }%
  \bibfield{journal}{%
  \Doi{10.1103/PhysRevB.81.125116}{\bibinfo {journal} {Phys. Rev. B}}\ }%
  \textbf{\bibinfo {volume} {81}},\ \bibinfo {pages} {125116} (\bibinfo {year}
  {2010})%
  \bibAnnoteFile{NoStop}{medvedeva-125116}%
\bibitem{schirmer-667}%
  \BibitemOpen
  \bibfield{author}{%
  \bibinfo {author} {\bibfnamefont{O.~F.}\ \bibnamefont{Schirmer}},\ }%
  \bibfield{journal}{%
  \bibinfo {journal} {J. Phys.: Condens. Matter}\ }%
  \textbf{\bibinfo {volume} {18}},\ \bibinfo {pages} {R667} (\bibinfo {year}
  {2006})%
  \bibAnnoteFile{NoStop}{schirmer-667}%
\bibitem{stoneham-255208}%
  \BibitemOpen
  \bibfield{author}{%
  \bibinfo {author} {\bibfnamefont{A.~M.}\ \bibnamefont{Stoneham}}, \bibinfo
  {author} {\bibfnamefont{J.}~\bibnamefont{Gavartin}}, \bibinfo {author}
  {\bibfnamefont{A.~L.}\ \bibnamefont{Shluger}}, \bibinfo {author}
  {\bibfnamefont{A.~V.}\ \bibnamefont{Kimmel}}, \bibinfo {author}
  {\bibfnamefont{D.~M.}\ \bibnamefont{Ramo}}, \bibinfo {author}
  {\bibfnamefont{H.~M.}\ \bibnamefont{Ronnow}}, \bibinfo {author}
  {\bibfnamefont{G.}~\bibnamefont{Aeppli}},\ and\ \bibinfo {author}
  {\bibfnamefont{C.}~\bibnamefont{Renner}},\ }%
  \bibfield{journal}{%
  \bibinfo {journal} {J. Phys.: Condens. Matter}\ }%
  \textbf{\bibinfo {volume} {19}},\ \bibinfo {pages} {255208} (\bibinfo {year}
  {2007})%
  \bibAnnoteFile{NoStop}{stoneham-255208}%
\bibitem{nie-066405}%
  \BibitemOpen
  \bibfield{author}{%
  \bibinfo {author} {\bibfnamefont{X.}~\bibnamefont{Nie}}, \bibinfo {author}
  {\bibfnamefont{S.-H.}\ \bibnamefont{Wei}},\ and\ \bibinfo {author}
  {\bibfnamefont{S.~B.}\ \bibnamefont{Zhang}},\ }%
  \bibfield{journal}{%
  \bibinfo {journal} {Phys. Rev. Lett.}\ }%
  \textbf{\bibinfo {volume} {88}},\ \bibinfo {pages} {066405} (\bibinfo {year}
  {2002})%
  \bibAnnoteFile{NoStop}{nie-066405}%
\bibitem{nie-075111}%
  \BibitemOpen
  \bibfield{author}{%
  \bibinfo {author} {\bibfnamefont{X.}~\bibnamefont{Nie}}, \bibinfo {author}
  {\bibfnamefont{S.-H.}\ \bibnamefont{Wei}},\ and\ \bibinfo {author}
  {\bibfnamefont{S.~B.}\ \bibnamefont{Zhang}},\ }%
  \bibfield{journal}{%
  \bibinfo {journal} {Phys. Rev. B}\ }%
  \textbf{\bibinfo {volume} {65}},\ \bibinfo {pages} {075111} (\bibinfo {year}
  {2002})%
  \bibAnnoteFile{NoStop}{nie-075111}%
\bibitem{kawazoe-939}%
  \BibitemOpen
  \bibfield{author}{%
  \bibinfo {author} {\bibfnamefont{H.}~\bibnamefont{Kawazoe}}, \bibinfo
  {author} {\bibfnamefont{M.}~\bibnamefont{Yasukawa}}, \bibinfo {author}
  {\bibfnamefont{H.}~\bibnamefont{Hyodo}}, \bibinfo {author}
  {\bibfnamefont{M.}~\bibnamefont{Kurita1}}, \bibinfo {author}
  {\bibfnamefont{H.}~\bibnamefont{Yanagi}},\ and\ \bibinfo {author}
  {\bibfnamefont{H.}~\bibnamefont{Hosono}},\ }%
  \bibfield{journal}{%
  \Doi{10.1038/40087}{\bibinfo {journal} {Nature}}\ }%
  \textbf{\bibinfo {volume} {389}},\ \bibinfo {pages} {939} (\bibinfo {year}
  {1997})%
  \bibAnnoteFile{NoStop}{kawazoe-939}%
\bibitem{saadi-272}%
  \BibitemOpen
  \bibfield{author}{%
  \bibinfo {author} {\bibfnamefont{S.}~\bibnamefont{Saadi}}, \bibinfo {author}
  {\bibfnamefont{A.}~\bibnamefont{Bouguelia}},\ and\ \bibinfo {author}
  {\bibfnamefont{M.}~\bibnamefont{Trari}},\ }%
  \bibfield{journal}{%
  \bibinfo {journal} {Solar Energy}\ }%
  \textbf{\bibinfo {volume} {80}},\ \bibinfo {pages} {272} (\bibinfo {year}
  {2006})%
  \bibAnnoteFile{NoStop}{saadi-272}%
\bibitem{scanlon-035101}%
  \BibitemOpen
  \bibfield{author}{%
  \bibinfo {author} {\bibfnamefont{D.~O.}\ \bibnamefont{Scanlon}}, \bibinfo
  {author} {\bibfnamefont{A.}~\bibnamefont{Walsh}}, \bibinfo {author}
  {\bibfnamefont{B.~J.}\ \bibnamefont{Morgan}}, \bibinfo {author}
  {\bibfnamefont{G.~W.}\ \bibnamefont{Watson}}, \bibinfo {author}
  {\bibfnamefont{D.~J.}\ \bibnamefont{Payne}},\ and\ \bibinfo {author}
  {\bibfnamefont{R.~G.}\ \bibnamefont{Egdell}},\ }%
  \bibfield{journal}{%
  \Doi{10.1103/PhysRevB.79.035101}{\bibinfo {journal} {Phys. Rev. B}}\ }%
  \textbf{\bibinfo {volume} {79}},\ \bibinfo {pages} {035101} (\bibinfo {year}
  {2009})%
  \bibAnnoteFile{NoStop}{scanlon-035101}%
\bibitem{arnold-075102}%
  \BibitemOpen
  \bibfield{author}{%
  \bibinfo {author} {\bibfnamefont{T.}~\bibnamefont{Arnold}}, \bibinfo {author}
  {\bibfnamefont{D.~J.}\ \bibnamefont{Payne}}, \bibinfo {author}
  {\bibfnamefont{A.}~\bibnamefont{Bourlange}}, \bibinfo {author}
  {\bibfnamefont{J.~P.}\ \bibnamefont{Hu}}, \bibinfo {author}
  {\bibfnamefont{R.~G.}\ \bibnamefont{Egdell}}, \bibinfo {author}
  {\bibfnamefont{L.~F.~J.}\ \bibnamefont{Piper}}, \bibinfo {author}
  {\bibfnamefont{L.}~\bibnamefont{Colakerol}}, \bibinfo {author}
  {\bibfnamefont{A.}~\bibnamefont{De~Masi}}, \bibinfo {author}
  {\bibfnamefont{P.-A.}\ \bibnamefont{Glans}}, \bibinfo {author}
  {\bibfnamefont{T.}~\bibnamefont{Learmonth}}, \bibinfo {author}
  {\bibfnamefont{K.~E.}\ \bibnamefont{Smith}}, \bibinfo {author}
  {\bibfnamefont{J.}~\bibnamefont{Guo}}, \bibinfo {author}
  {\bibfnamefont{D.~O.}\ \bibnamefont{Scanlon}}, \bibinfo {author}
  {\bibfnamefont{A.}~\bibnamefont{Walsh}}, \bibinfo {author}
  {\bibfnamefont{B.~J.}\ \bibnamefont{Morgan}},\ and\ \bibinfo {author}
  {\bibfnamefont{G.~W.}\ \bibnamefont{Watson}},\ }%
  \bibfield{journal}{%
  \Doi{10.1103/PhysRevB.79.075102}{\bibinfo {journal} {Phys. Rev. B}}\ }%
  \textbf{\bibinfo {volume} {79}},\ \bibinfo {pages} {075102} (\bibinfo {year}
  {2009})%
  \bibAnnoteFile{NoStop}{arnold-075102}%
\bibitem{huda-035205}%
  \BibitemOpen
  \bibfield{author}{%
  \bibinfo {author} {\bibfnamefont{M.~N.}\ \bibnamefont{Huda}}, \bibinfo
  {author} {\bibfnamefont{Y.}~\bibnamefont{Yan}}, \bibinfo {author}
  {\bibfnamefont{A.}~\bibnamefont{Walsh}}, \bibinfo {author}
  {\bibfnamefont{S.-H.}\ \bibnamefont{Wei}},\ and\ \bibinfo {author}
  {\bibfnamefont{M.~M.}\ \bibnamefont{Al-Jassim}},\ }%
  \bibfield{journal}{%
  \Doi{10.1103/PhysRevB.80.035205}{\bibinfo {journal} {Phys. Rev. B}}\ }%
  \textbf{\bibinfo {volume} {80}},\ \bibinfo {pages} {035205} (\bibinfo {year}
  {2009})%
  \bibAnnoteFile{NoStop}{huda-035205}%
\bibitem{scanlon-4568}%
  \BibitemOpen
  \bibfield{author}{%
  \bibinfo {author} {\bibfnamefont{D.~O.}\ \bibnamefont{Scanlon}}, \bibinfo
  {author} {\bibfnamefont{G.~W.}\ \bibnamefont{Watson}},\ and\ \bibinfo
  {author} {\bibfnamefont{A.}~\bibnamefont{Walsh}},\ }%
  \bibfield{journal}{%
  \bibinfo {journal} {Chem. Mater.}\ }%
  \textbf{\bibinfo {volume} {21}},\ \bibinfo {pages} {4568} (\bibinfo {year}
  {2009})%
  \bibAnnoteFile{NoStop}{scanlon-4568}%
\bibitem{kudo-220}%
  \BibitemOpen
  \bibfield{author}{%
  \bibinfo {author} {\bibfnamefont{A.}~\bibnamefont{Kudo}}, \bibinfo {author}
  {\bibfnamefont{H.}~\bibnamefont{Yanagi}}, \bibinfo {author}
  {\bibfnamefont{H.}~\bibnamefont{Hosono}},\ and\ \bibinfo {author}
  {\bibfnamefont{H.}~\bibnamefont{Kawazoe}},\ }%
  \bibfield{journal}{%
  \Doi{10.1063/1.121761}{\bibinfo {journal} {Appl. Phys. Lett.}}\ }%
  \textbf{\bibinfo {volume} {73}},\ \bibinfo {pages} {220} (\bibinfo {year}
  {1998})%
  \bibAnnoteFile{NoStop}{kudo-220}%
\bibitem{godinho-2798}%
  \BibitemOpen
  \bibfield{author}{%
  \bibinfo {author} {\bibfnamefont{K.~G.}\ \bibnamefont{Godinho}}, \bibinfo
  {author} {\bibfnamefont{G.~W.}\ \bibnamefont{Watson}}, \bibinfo {author}
  {\bibfnamefont{A.}~\bibnamefont{Walsh}}, \bibinfo {author}
  {\bibfnamefont{A.}~\bibnamefont{Green}}, \bibinfo {author}
  {\bibfnamefont{J.}~\bibnamefont{Harmer}}, \bibinfo {author}
  {\bibfnamefont{D.}~\bibnamefont{Payne}},\ and\ \bibinfo {author}
  {\bibfnamefont{R.}~\bibnamefont{Egdell}},\ }%
  \bibfield{journal}{%
  \bibinfo {journal} {J. Mater. Chem.}\ }%
  \textbf{\bibinfo {volume} {18}},\ \bibinfo {pages} {2798} (\bibinfo {year}
  {2008})%
  \bibAnnoteFile{NoStop}{godinho-2798}%
\bibitem{scanlon-096405}%
  \BibitemOpen
  \bibfield{author}{%
  \bibinfo {author} {\bibfnamefont{D.~O.}\ \bibnamefont{Scanlon}}, \bibinfo
  {author} {\bibfnamefont{B.~J.}\ \bibnamefont{Morgan}}, \bibinfo {author}
  {\bibfnamefont{G.~W.}\ \bibnamefont{Watson}},\ and\ \bibinfo {author}
  {\bibfnamefont{A.}~\bibnamefont{Walsh}},\ }%
  \bibfield{journal}{%
  \bibinfo {journal} {Phys. Rev. Lett.}\ }%
  \textbf{\bibinfo {volume} {103}},\ \bibinfo {pages} {096405} (\bibinfo {year}
  {2009})%
  \bibAnnoteFile{NoStop}{scanlon-096405}%
\bibitem{minami-971}%
  \BibitemOpen
  \bibfield{author}{%
  \bibinfo {author} {\bibfnamefont{T.}~\bibnamefont{Minami}}, \bibinfo {author}
  {\bibfnamefont{H.}~\bibnamefont{Sonohara}}, \bibinfo {author}
  {\bibfnamefont{T.}~\bibnamefont{Kakumu}},\ and\ \bibinfo {author}
  {\bibfnamefont{S.}~\bibnamefont{Takata}},\ }%
  \bibfield{journal}{%
  \bibinfo {journal} {Jpn. J. Appl. Phys.}\ }%
  \textbf{\bibinfo {volume} {34}},\ \bibinfo {pages} {L971} (\bibinfo {year}
  {1995})%
  \bibAnnoteFile{NoStop}{minami-971}%
\bibitem{taylor-90}%
  \BibitemOpen
  \bibfield{author}{%
  \bibinfo {author} {\bibfnamefont{M.~P.}\ \bibnamefont{Taylor}}, \bibinfo
  {author} {\bibfnamefont{D.~W.}\ \bibnamefont{Readey}}, \bibinfo {author}
  {\bibfnamefont{C.~W.}\ \bibnamefont{Teplin}}, \bibinfo {author}
  {\bibfnamefont{M.~F. A.~M.}\ \bibnamefont{van Hest}}, \bibinfo {author}
  {\bibfnamefont{J.~L.}\ \bibnamefont{Alleman}}, \bibinfo {author}
  {\bibfnamefont{M.~S.}\ \bibnamefont{Dabney}}, \bibinfo {author}
  {\bibfnamefont{L.~M.}\ \bibnamefont{Gedvilas}}, \bibinfo {author}
  {\bibfnamefont{B.~M.}\ \bibnamefont{Keyes}}, \bibinfo {author}
  {\bibfnamefont{B.}~\bibnamefont{To}}, \bibinfo {author}
  {\bibfnamefont{J.~D.}\ \bibnamefont{Perkins}},\ and\ \bibinfo {author}
  {\bibfnamefont{D.~S.}\ \bibnamefont{Ginley}},\ }%
  \bibfield{journal}{%
  \bibinfo {journal} {Meas. Sci. Technol.}\ }%
  \textbf{\bibinfo {volume} {16}},\ \bibinfo {pages} {90} (\bibinfo {year}
  {2005})%
  \bibAnnoteFile{NoStop}{taylor-90}%
\bibitem{hiramatsu-3033}%
  \BibitemOpen
  \bibfield{author}{%
  \bibinfo {author} {\bibfnamefont{H.}~\bibnamefont{Hiramatsu}}, \bibinfo
  {author} {\bibfnamefont{W.-S.}\ \bibnamefont{Seo}},\ and\ \bibinfo {author}
  {\bibfnamefont{K.}~\bibnamefont{Koumoto}},\ }%
  \bibfield{journal}{%
  \bibinfo {journal} {Chem. Mater.}\ }%
  \textbf{\bibinfo {volume} {10}},\ \bibinfo {pages} {3033} (\bibinfo {year}
  {1998})%
  \bibAnnoteFile{NoStop}{hiramatsu-3033}%
\bibitem{kumar-073703}%
  \BibitemOpen
  \bibfield{author}{%
  \bibinfo {author} {\bibfnamefont{B.}~\bibnamefont{Kumar}}, \bibinfo {author}
  {\bibfnamefont{H.}~\bibnamefont{Gong}},\ and\ \bibinfo {author}
  {\bibfnamefont{R.}~\bibnamefont{Akkipeddi}},\ }%
  \bibfield{journal}{%
  \bibinfo {journal} {J. Appl. Phys.}\ }%
  \textbf{\bibinfo {volume} {98}},\ \bibinfo {pages} {073703} (\bibinfo {year}
  {2005})%
  \bibAnnoteFile{NoStop}{kumar-073703}%
\bibitem{leenheer-115215}%
  \BibitemOpen
  \bibfield{author}{%
  \bibinfo {author} {\bibfnamefont{A.~J.}\ \bibnamefont{Leenheer}}, \bibinfo
  {author} {\bibfnamefont{J.~D.}\ \bibnamefont{Perkins}}, \bibinfo {author}
  {\bibfnamefont{M.~F. A.~M.}\ \bibnamefont{van Hest}}, \bibinfo {author}
  {\bibfnamefont{J.~J.}\ \bibnamefont{Berry}}, \bibinfo {author}
  {\bibfnamefont{R.~P.}\ \bibnamefont{O'Hayre}},\ and\ \bibinfo {author}
  {\bibfnamefont{D.~S.}\ \bibnamefont{Ginley}},\ }%
  \bibfield{journal}{%
  \bibinfo {journal} {Phys. Rev. B}\ }%
  \textbf{\bibinfo {volume} {77}},\ \bibinfo {pages} {115215} (\bibinfo {year}
  {2008})%
  \bibAnnoteFile{NoStop}{leenheer-115215}%
\bibitem{moriga-312}%
  \BibitemOpen
  \bibfield{author}{%
  \bibinfo {author} {\bibfnamefont{T.}~\bibnamefont{Moriga}}, \bibinfo {author}
  {\bibfnamefont{T.}~\bibnamefont{Okamoto}}, \bibinfo {author}
  {\bibfnamefont{K.}~\bibnamefont{Hiruta}}, \bibinfo {author}
  {\bibfnamefont{A.}~\bibnamefont{Fujiwara}},\ and\ \bibinfo {author}
  {\bibfnamefont{I.}~\bibnamefont{Nakabayashi}},\ }%
  \bibfield{journal}{%
  \bibinfo {journal} {J. Solid State Chem.}\ }%
  \textbf{\bibinfo {volume} {155}},\ \bibinfo {pages} {312} (\bibinfo {year}
  {2000})%
  \bibAnnoteFile{NoStop}{moriga-312}%
\bibitem{nomura-488}%
  \BibitemOpen
  \bibfield{author}{%
  \bibinfo {author} {\bibfnamefont{K.}~\bibnamefont{Nomura}}, \bibinfo {author}
  {\bibfnamefont{H.}~\bibnamefont{Ohta}}, \bibinfo {author}
  {\bibfnamefont{A.}~\bibnamefont{Takagi}}, \bibinfo {author}
  {\bibfnamefont{T.}~\bibnamefont{Kamiya}}, \bibinfo {author}
  {\bibfnamefont{M.}~\bibnamefont{Hirano}},\ and\ \bibinfo {author}
  {\bibfnamefont{H.}~\bibnamefont{Hosono}},\ }%
  \bibfield{journal}{%
  \bibinfo {journal} {Nature}\ }%
  \textbf{\bibinfo {volume} {432}},\ \bibinfo {pages} {488} (\bibinfo {year}
  {2004})%
  \bibAnnoteFile{NoStop}{nomura-488}%
\bibitem{sun-1897}%
  \BibitemOpen
  \bibfield{author}{%
  \bibinfo {author} {\bibfnamefont{Y.}~\bibnamefont{Sun}}\ and\ \bibinfo
  {author} {\bibfnamefont{J.}~\bibnamefont{Rogers}},\ }%
  \bibfield{journal}{%
  \bibinfo {journal} {Adv. Mater.}\ }%
  \textbf{\bibinfo {volume} {19}},\ \bibinfo {pages} {1897} (\bibinfo {year}
  {2007})%
  \bibAnnoteFile{NoStop}{sun-1897}%
\bibitem{lee-843}%
  \BibitemOpen
  \bibfield{author}{%
  \bibinfo {author} {\bibfnamefont{D.-H.}\ \bibnamefont{Lee}}, \bibinfo
  {author} {\bibfnamefont{Y.-J.}\ \bibnamefont{Chang}}, \bibinfo {author}
  {\bibfnamefont{G.}~\bibnamefont{Herman}},\ and\ \bibinfo {author}
  {\bibfnamefont{C.-H.}\ \bibnamefont{Chang}},\ }%
  \bibfield{journal}{%
  \bibinfo {journal} {Adv. Mater.}\ }%
  \textbf{\bibinfo {volume} {19}},\ \bibinfo {pages} {843} (\bibinfo {year}
  {2008})%
  \bibAnnoteFile{NoStop}{lee-843}%
\bibitem{taylor-3169}%
  \BibitemOpen
  \bibfield{author}{%
  \bibinfo {author} {\bibfnamefont{M.~P.}\ \bibnamefont{Taylor}}, \bibinfo
  {author} {\bibfnamefont{D.~W.}\ \bibnamefont{Readey}}, \bibinfo {author}
  {\bibfnamefont{M.~F. A.~M.}\ \bibnamefont{van Hest}}, \bibinfo {author}
  {\bibfnamefont{C.~W.}\ \bibnamefont{Teplin}}, \bibinfo {author}
  {\bibfnamefont{J.~L.}\ \bibnamefont{Alleman}}, \bibinfo {author}
  {\bibfnamefont{M.~S.}\ \bibnamefont{Dabney}}, \bibinfo {author}
  {\bibfnamefont{L.~M.}\ \bibnamefont{Gedvilas}}, \bibinfo {author}
  {\bibfnamefont{B.~M.}\ \bibnamefont{Keyes}}, \bibinfo {author}
  {\bibfnamefont{B.}~\bibnamefont{To}}, \bibinfo {author}
  {\bibfnamefont{J.~D.}\ \bibnamefont{Perkins}},\ and\ \bibinfo {author}
  {\bibfnamefont{D.~S.}\ \bibnamefont{Ginley}},\ }%
  \bibfield{journal}{%
  \bibinfo {journal} {Adv. Funct. Mater.}\ }%
  \textbf{\bibinfo {volume} {18}},\ \bibinfo {pages} {3169} (\bibinfo {year}
  {2008})%
  \bibAnnoteFile{NoStop}{taylor-3169}%
\bibitem{mott-1987}%
  \BibitemOpen
  \bibfield{author}{%
  \bibinfo {author} {\bibfnamefont{N.~F.}\ \bibnamefont{Mott}},\ }%
  \emph{\bibinfo {title} {Conduction in non-crystalline materials}},\ \bibinfo
  {edition} {1st}\ ed.\ (\bibinfo {publisher} {Oxford Science Publications},\
  \bibinfo {address} {Oxford},\ \bibinfo {year} {1987})%
  \bibAnnoteFile{NoStop}{mott-1987}%
\bibitem{walsh-5119}%
  \BibitemOpen
  \bibfield{author}{%
  \bibinfo {author} {\bibfnamefont{A.}~\bibnamefont{Walsh}}, \bibinfo {author}
  {\bibfnamefont{J.~L.~F.}\ \bibnamefont{DaSilva}},\ and\ \bibinfo {author}
  {\bibfnamefont{S.-H.}\ \bibnamefont{Wei}},\ }%
  \bibfield{journal}{%
  \bibinfo {journal} {Chem. Mater.}\ }%
  \textbf{\bibinfo {volume} {21}},\ \bibinfo {pages} {5119} (\bibinfo {year}
  {2009})%
  \bibAnnoteFile{NoStop}{walsh-5119}%
\bibitem{robertson-1026}%
  \BibitemOpen
  \bibfield{author}{%
  \bibinfo {author} {\bibfnamefont{J.}~\bibnamefont{Robertson}},\ }%
  \bibfield{journal}{%
  \bibinfo {journal} {Phys. Status Solidi B}\ }%
  \textbf{\bibinfo {volume} {245}},\ \bibinfo {pages} {1026} (\bibinfo {year}
  {2008})%
  \bibAnnoteFile{NoStop}{robertson-1026}%
\bibitem{hosono-2796}%
  \BibitemOpen
  \bibfield{author}{%
  \bibinfo {author} {\bibfnamefont{H.}~\bibnamefont{Hosono}}, \bibinfo {author}
  {\bibfnamefont{K.}~\bibnamefont{Nomura}}, \bibinfo {author}
  {\bibfnamefont{Y.}~\bibnamefont{Ogo}}, \bibinfo {author}
  {\bibfnamefont{T.}~\bibnamefont{Uruga}},\ and\ \bibinfo {author}
  {\bibfnamefont{T.}~\bibnamefont{Kamiya}},\ }%
  \bibfield{journal}{%
  \bibinfo {journal} {J. Non-Cryst. Solids}\ }%
  \textbf{\bibinfo {volume} {354}},\ \bibinfo {pages} {2796} (\bibinfo {year}
  {2008})%
  \bibAnnoteFile{NoStop}{hosono-2796}%
\bibitem{hosono-6000}%
  \BibitemOpen
  \bibfield{author}{%
  \bibinfo {author} {\bibfnamefont{H.}~\bibnamefont{Hosono}},\ }%
  \bibfield{journal}{%
  \bibinfo {journal} {Thin Solid Films}\ }%
  \textbf{\bibinfo {volume} {515}},\ \bibinfo {pages} {6000} (\bibinfo {year}
  {2007})%
  \bibAnnoteFile{NoStop}{hosono-6000}%
\bibitem{fortunato-242}%
  \BibitemOpen
  \bibfield{author}{%
  \bibinfo {author} {\bibfnamefont{E.}~\bibnamefont{Fortunato}}, \bibinfo
  {author} {\bibfnamefont{D.}~\bibnamefont{Ginley}}, \bibinfo {author}
  {\bibfnamefont{H.}~\bibnamefont{Hosono}},\ and\ \bibinfo {author}
  {\bibfnamefont{D.~C.}\ \bibnamefont{Paine}},\ }%
  \bibfield{journal}{%
  \bibinfo {journal} {MRS Bull.}\ }%
  \textbf{\bibinfo {volume} {32}},\ \bibinfo {pages} {242} (\bibinfo {year}
  {2007})%
  \bibAnnoteFile{NoStop}{fortunato-242}%
\bibitem{hoel-3569}%
  \BibitemOpen
  \bibfield{author}{%
  \bibinfo {author} {\bibfnamefont{C.~A.}\ \bibnamefont{Hoel}}, \bibinfo
  {author} {\bibfnamefont{T.~O.}\ \bibnamefont{Mason}}, \bibinfo {author}
  {\bibfnamefont{J.-F.}\ \bibnamefont{Gaillard}},\ and\ \bibinfo {author}
  {\bibfnamefont{K.~R.}\ \bibnamefont{Poeppelmeier}},\ }%
  \bibfield{journal}{%
  \bibinfo {journal} {Chem. Mater.}\ }%
  \textbf{\bibinfo {volume} {22}},\ \bibinfo {pages} {3569} (\bibinfo {year}
  {2010})%
  \bibAnnoteFile{NoStop}{hoel-3569}%
\bibitem{catlow-2234}%
  \BibitemOpen
  \bibfield{author}{%
  \bibinfo {author} {\bibfnamefont{C.~R.~A.}\ \bibnamefont{Catlow}}, \bibinfo
  {author} {\bibfnamefont{S.}~\bibnamefont{French}}, \bibinfo {author}
  {\bibfnamefont{A.~A.}\ \bibnamefont{Sokol}}, \bibinfo {author}
  {\bibfnamefont{A.}~\bibnamefont{Al-Sunaidi}},\ and\ \bibinfo {author}
  {\bibfnamefont{S.}~\bibnamefont{Woodley}},\ }%
  \bibfield{journal}{%
  \bibinfo {journal} {J. Comp. Chem.}\ }%
  \textbf{\bibinfo {volume} {29}},\ \bibinfo {pages} {2234} (\bibinfo {year}
  {2008})%
  \bibAnnoteFile{NoStop}{catlow-2234}%
\bibitem{lany-235104}%
  \BibitemOpen
  \bibfield{author}{%
  \bibinfo {author} {\bibfnamefont{S.}~\bibnamefont{Lany}}\ and\ \bibinfo
  {author} {\bibfnamefont{A.}~\bibnamefont{Zunger}},\ }%
  \bibfield{journal}{%
  \bibinfo {journal} {Phys. Rev. B}\ }%
  \textbf{\bibinfo {volume} {78}},\ \bibinfo {pages} {235104} (\bibinfo {year}
  {2008})%
  \bibAnnoteFile{NoStop}{lany-235104}%
\bibitem{janotti-165202}%
  \BibitemOpen
  \bibfield{author}{%
  \bibinfo {author} {\bibfnamefont{A.}~\bibnamefont{Janotti}}\ and\ \bibinfo
  {author} {\bibfnamefont{C.~G.}\ \bibnamefont{Van~de Walle}},\ }%
  \bibfield{journal}{%
  \Doi{10.1103/PhysRevB.76.165202}{\bibinfo {journal} {Phys. Rev. B}}\ }%
  \textbf{\bibinfo {volume} {76}},\ \bibinfo {pages} {165202} (\bibinfo {year}
  {2007})%
  \bibAnnoteFile{NoStop}{janotti-165202}%
\bibitem{dft1}%
  \BibitemOpen
  \bibfield{author}{%
  \bibinfo {author} {\bibfnamefont{W.}~\bibnamefont{Kohn}}\ and\ \bibinfo
  {author} {\bibfnamefont{L.~J.}\ \bibnamefont{Sham}},\ }%
  \bibfield{journal}{%
  \bibinfo {journal} {Phys. Rev.}\ }%
  \textbf{\bibinfo {volume} {140}},\ \bibinfo {pages} {A1133} (\bibinfo {year}
  {1965})%
  \bibAnnoteFile{NoStop}{dft1}%
\bibitem{dft2}%
  \BibitemOpen
  \bibfield{author}{%
  \bibinfo {author} {\bibfnamefont{P.}~\bibnamefont{Hohenberg}}\ and\ \bibinfo
  {author} {\bibfnamefont{W.}~\bibnamefont{Kohn}},\ }%
  \bibfield{journal}{%
  \bibinfo {journal} {Phys. Rev.}\ }%
  \textbf{\bibinfo {volume} {136}},\ \bibinfo {pages} {B864} (\bibinfo {year}
  {1964})%
  \bibAnnoteFile{NoStop}{dft2}%
\bibitem{vasp1}%
  \BibitemOpen
  \bibfield{author}{%
  \bibinfo {author} {\bibfnamefont{G.}~\bibnamefont{Kresse}}\ and\ \bibinfo
  {author} {\bibfnamefont{J.}~\bibnamefont{Furthm\"uller}},\ }%
  \bibfield{journal}{%
  \bibinfo {journal} {Phys. Rev. B}\ }%
  \textbf{\bibinfo {volume} {54}},\ \bibinfo {pages} {11169} (\bibinfo {year}
  {1996})%
  \bibAnnoteFile{NoStop}{vasp1}%
\bibitem{castep}%
  \BibitemOpen
  \bibfield{author}{%
  \bibinfo {author} {\bibfnamefont{M.}~\bibnamefont{Segall}}, \bibinfo {author}
  {\bibfnamefont{P.}~\bibnamefont{Lindan}}, \bibinfo {author}
  {\bibfnamefont{M.}~\bibnamefont{Probert}}, \bibinfo {author}
  {\bibfnamefont{C.}~\bibnamefont{Pickard}}, \bibinfo {author}
  {\bibfnamefont{P.}~\bibnamefont{Hasnip}}, \bibinfo {author}
  {\bibfnamefont{S.}~\bibnamefont{Clark}},\ and\ \bibinfo {author}
  {\bibfnamefont{M.}~\bibnamefont{Payne}},\ }%
  \bibfield{journal}{%
  \bibinfo {journal} {J. Phys.: Cond. Matter}\ }%
  \textbf{\bibinfo {volume} {14}},\ \bibinfo {pages} {2717} (\bibinfo {year}
  {2002})%
  \bibAnnoteFile{NoStop}{castep}%
\bibitem{espresso}%
  \BibitemOpen
  \bibfield{author}{%
  \bibinfo {author} {\bibfnamefont{P.}~\bibnamefont{Giannozzi}}, \bibinfo
  {author} {\bibfnamefont{S.}~\bibnamefont{Baroni}}, \bibinfo {author}
  {\bibfnamefont{N.}~\bibnamefont{Bonini}}, \bibinfo {author}
  {\bibfnamefont{M.}~\bibnamefont{Calandra}}, \bibinfo {author}
  {\bibfnamefont{R.}~\bibnamefont{Car}}, \bibinfo {author}
  {\bibfnamefont{C.}~\bibnamefont{Cavazzoni}}, \bibinfo {author}
  {\bibfnamefont{D.}~\bibnamefont{Ceresoli}}, \bibinfo {author}
  {\bibfnamefont{G.}~\bibnamefont{Chiarotti}}, \bibinfo {author}
  {\bibfnamefont{M.}~\bibnamefont{Cococcioni}}, \bibinfo {author}
  {\bibfnamefont{I.}~\bibnamefont{Dabo}}, \emph{et~al.},\ }%
  \bibfield{journal}{%
  \bibinfo {journal} {J. Phys.: Cond. Matter}\ }%
  \textbf{\bibinfo {volume} {21}},\ \bibinfo {pages} {395502} (\bibinfo {year}
  {2009})%
  \bibAnnoteFile{NoStop}{espresso}%
\bibitem{aims1}%
  \BibitemOpen
  \bibfield{author}{%
  \bibinfo {author} {\bibfnamefont{V.}~\bibnamefont{Blum}}, \bibinfo {author}
  {\bibfnamefont{R.}~\bibnamefont{Gehrke}}, \bibinfo {author}
  {\bibfnamefont{F.}~\bibnamefont{Hanke}}, \bibinfo {author}
  {\bibfnamefont{P.}~\bibnamefont{Havu}}, \bibinfo {author}
  {\bibfnamefont{V.}~\bibnamefont{Havu}}, \bibinfo {author}
  {\bibfnamefont{X.}~\bibnamefont{Ren}}, \bibinfo {author}
  {\bibfnamefont{K.}~\bibnamefont{Reuter}},\ and\ \bibinfo {author}
  {\bibfnamefont{M.}~\bibnamefont{Scheffler}},\ }%
  \bibfield{journal}{%
  \Doi{DOI: 10.1016/j.cpc.2009.06.022}{\bibinfo {journal} {Comp. Phys. Comm.}}\
  }%
  \textbf{\bibinfo {volume} {180}},\ \bibinfo {pages} {2175} (\bibinfo {year}
  {2009})%
  \bibAnnoteFile{NoStop}{aims1}%
\bibitem{martin-2004}%
  \BibitemOpen
  \bibfield{author}{%
  \bibinfo {author} {\bibfnamefont{R.~M.}\ \bibnamefont{Martin}},\ }%
  \emph{\bibinfo {title} {Electronic Structure}},\ \bibinfo {edition} {1st}\
  ed.\ (\bibinfo {publisher} {Cambridge University Press},\ \bibinfo {address}
  {Cambridge},\ \bibinfo {year} {2004})%
  \bibAnnoteFile{NoStop}{martin-2004}%
\bibitem{payne-1045}%
  \BibitemOpen
  \bibfield{author}{%
  \bibinfo {author} {\bibfnamefont{M.~C.}\ \bibnamefont{Payne}}, \bibinfo
  {author} {\bibfnamefont{M.~P.}\ \bibnamefont{Teter}}, \bibinfo {author}
  {\bibfnamefont{D.~C.}\ \bibnamefont{Allan}}, \bibinfo {author}
  {\bibfnamefont{T.~A.}\ \bibnamefont{Arias}},\ and\ \bibinfo {author}
  {\bibfnamefont{J.~D.}\ \bibnamefont{Joannopoulos}},\ }%
  \bibfield{journal}{%
  \Doi{10.1103/RevModPhys.64.1045}{\bibinfo {journal} {Rev. Mod. Phys.}}\ }%
  \textbf{\bibinfo {volume} {64}},\ \bibinfo {pages} {1045} (\bibinfo {year}
  {1992})%
  \bibAnnoteFile{NoStop}{payne-1045}%
\bibitem{woodley-937}%
  \BibitemOpen
  \bibfield{author}{%
  \bibinfo {author} {\bibfnamefont{S.~M.}\ \bibnamefont{Woodley}}\ and\
  \bibinfo {author} {\bibfnamefont{C.~R.~A.}\ \bibnamefont{Catlow}},\ }%
  \bibfield{journal}{%
  \bibinfo {journal} {Nat. Mater.}\ }%
  \textbf{\bibinfo {volume} {7}},\ \bibinfo {pages} {937} (\bibinfo {year}
  {2008})%
  \bibAnnoteFile{NoStop}{woodley-937}%
\bibitem{walsh-8446}%
  \BibitemOpen
  \bibfield{author}{%
  \bibinfo {author} {\bibfnamefont{A.}~\bibnamefont{Walsh}}\ and\ \bibinfo
  {author} {\bibfnamefont{S.~M.}\ \bibnamefont{Woodley}},\ }%
  \bibfield{journal}{%
  \bibinfo {journal} {Phys. Chem. Chem. Phys.}\ }%
  \textbf{\bibinfo {volume} {12}},\ \bibinfo {pages} {8446} (\bibinfo {year}
  {2010})%
  \bibAnnoteFile{NoStop}{walsh-8446}%
\bibitem{trimarchi-295212}%
  \BibitemOpen
  \bibfield{author}{%
  \bibinfo {author} {\bibfnamefont{G.}~\bibnamefont{Trimarchi}}\ and\ \bibinfo
  {author} {\bibfnamefont{A.}~\bibnamefont{Zunger}},\ }%
  \bibfield{journal}{%
  \bibinfo {journal} {J. Phys.: Condens. Matter}\ }%
  \textbf{\bibinfo {volume} {20}},\ \bibinfo {pages} {295212} (\bibinfo {year}
  {2008})%
  \bibAnnoteFile{NoStop}{trimarchi-295212}%
\bibitem{hautier-3762}%
  \BibitemOpen
  \bibfield{author}{%
  \bibinfo {author} {\bibfnamefont{G.}~\bibnamefont{Hautier}}, \bibinfo
  {author} {\bibfnamefont{C.}~\bibnamefont{Fischer}}, \bibinfo {author}
  {\bibfnamefont{A.}~\bibnamefont{Jain}}, \bibinfo {author}
  {\bibfnamefont{T.}~\bibnamefont{Mueller}},\ and\ \bibinfo {author}
  {\bibfnamefont{G.}~\bibnamefont{Ceder}},\ }%
  \bibfield{journal}{%
  \bibinfo {journal} {Chem. Mater.}\ }%
  \textbf{\bibinfo {volume} {22}},\ \bibinfo {pages} {3762} (\bibinfo {year}
  {2010})%
  \bibAnnoteFile{NoStop}{hautier-3762}%
\bibitem{lda}%
  \BibitemOpen
  \bibfield{author}{%
  \bibinfo {author} {\bibfnamefont{D.~M.}\ \bibnamefont{Ceperley}}\ and\
  \bibinfo {author} {\bibfnamefont{B.~J.}\ \bibnamefont{Alder}},\ }%
  \bibfield{journal}{%
  \bibinfo {journal} {Phys. Rev. Lett.}\ }%
  \textbf{\bibinfo {volume} {45}},\ \bibinfo {pages} {566} (\bibinfo {year}
  {1980})%
  \bibAnnoteFile{NoStop}{lda}%
\bibitem{pbe}%
  \BibitemOpen
  \bibfield{author}{%
  \bibinfo {author} {\bibfnamefont{J.~P.}\ \bibnamefont{Perdew}}, \bibinfo
  {author} {\bibfnamefont{K.}~\bibnamefont{Burke}},\ and\ \bibinfo {author}
  {\bibfnamefont{M.}~\bibnamefont{Ernzerhof}},\ }%
  \bibfield{journal}{%
  \bibinfo {journal} {Phys. Rev. Lett.}\ }%
  \textbf{\bibinfo {volume} {77}},\ \bibinfo {pages} {3865} (\bibinfo {year}
  {1996})%
  \bibAnnoteFile{NoStop}{pbe}%
\bibitem{perdew-5048}%
  \BibitemOpen
  \bibfield{author}{%
  \bibinfo {author} {\bibfnamefont{J.~P.}\ \bibnamefont{Perdew}}\ and\ \bibinfo
  {author} {\bibfnamefont{A.}~\bibnamefont{Zunger}},\ }%
  \bibfield{journal}{%
  \bibinfo {journal} {Phys. Rev. B}\ }%
  \textbf{\bibinfo {volume} {23}},\ \bibinfo {pages} {5048} (\bibinfo {year}
  {1981})%
  \bibAnnoteFile{NoStop}{perdew-5048}%
\bibitem{perdew-1884}%
  \BibitemOpen
  \bibfield{author}{%
  \bibinfo {author} {\bibfnamefont{J.~P.}\ \bibnamefont{Perdew}}\ and\ \bibinfo
  {author} {\bibfnamefont{M.}~\bibnamefont{Levy}},\ }%
  \bibfield{journal}{%
  \bibinfo {journal} {Phys. Rev. Lett.}\ }%
  \textbf{\bibinfo {volume} {51}},\ \bibinfo {pages} {1884} (\bibinfo {year}
  {1983})%
  \bibAnnoteFile{NoStop}{perdew-1884}%
\bibitem{sham-1888}%
  \BibitemOpen
  \bibfield{author}{%
  \bibinfo {author} {\bibfnamefont{L.~J.}\ \bibnamefont{Sham}}\ and\ \bibinfo
  {author} {\bibfnamefont{M.}~\bibnamefont{Schl\"uter}},\ }%
  \bibfield{journal}{%
  \bibinfo {journal} {Phys. Rev. Lett.}\ }%
  \textbf{\bibinfo {volume} {51}},\ \bibinfo {pages} {1888} (\bibinfo {year}
  {1983})%
  \bibAnnoteFile{NoStop}{sham-1888}%
\bibitem{hedin-796}%
  \BibitemOpen
  \bibfield{author}{%
  \bibinfo {author} {\bibfnamefont{L.}~\bibnamefont{Hedin}},\ }%
  \bibfield{journal}{%
  \Doi{10.1103/PhysRev.139.A796}{\bibinfo {journal} {Phys. Rev.}}\ }%
  \textbf{\bibinfo {volume} {139}},\ \bibinfo {pages} {A796} (\bibinfo {year}
  {1965})%
  \bibAnnoteFile{NoStop}{hedin-796}%
\bibitem{cora-2004}%
  \BibitemOpen
  \bibfield{author}{%
  \bibinfo {author} {\bibfnamefont{F.}~\bibnamefont{Cora}}, \bibinfo {author}
  {\bibfnamefont{M.}~\bibnamefont{Alfredsson}}, \bibinfo {author}
  {\bibfnamefont{G.}~\bibnamefont{Mallia}}, \bibinfo {author}
  {\bibfnamefont{D.~S.}\ \bibnamefont{Middlemiss}}, \bibinfo {author}
  {\bibfnamefont{W.~C.}\ \bibnamefont{Mackrodt}}, \bibinfo {author}
  {\bibfnamefont{R.}~\bibnamefont{Dovesi}},\ and\ \bibinfo {author}
  {\bibfnamefont{R.}~\bibnamefont{Orlando}},\ }%
  \emph{\bibinfo {title} {''The Performance of Hybrid Density Functionals in
  Solid State Chemistry'' in Structure and Bonding}},\ \bibinfo {edition}
  {1st}\ ed.\ (\bibinfo {publisher} {Springer},\ \bibinfo {address} {Berlin},\
  \bibinfo {year} {2004})%
  \bibAnnoteFile{NoStop}{cora-2004}%
\bibitem{hse}%
  \BibitemOpen
  \bibfield{author}{%
  \bibinfo {author} {\bibfnamefont{J.}~\bibnamefont{Heyd}}, \bibinfo {author}
  {\bibfnamefont{G.~E.}\ \bibnamefont{Scuseria}},\ and\ \bibinfo {author}
  {\bibfnamefont{M.}~\bibnamefont{Ernzerhof}},\ }%
  \bibfield{journal}{%
  \Doi{10.1063/1.1564060}{\bibinfo {journal} {J. Chem. Phys.}}\ }%
  \textbf{\bibinfo {volume} {118}},\ \bibinfo {pages} {8207} (\bibinfo {year}
  {2003})%
  \bibAnnoteFile{NoStop}{hse}%
\bibitem{paier-154709}%
  \BibitemOpen
  \bibfield{author}{%
  \bibinfo {author} {\bibfnamefont{J.}~\bibnamefont{Paier}}, \bibinfo {author}
  {\bibfnamefont{M.}~\bibnamefont{Marsman}}, \bibinfo {author}
  {\bibfnamefont{K.}~\bibnamefont{Hummer}}, \bibinfo {author}
  {\bibfnamefont{G.}~\bibnamefont{Kresse}}, \bibinfo {author}
  {\bibfnamefont{I.~C.}\ \bibnamefont{Gerber}},\ and\ \bibinfo {author}
  {\bibfnamefont{J.~G.}\ \bibnamefont{Angyan}},\ }%
  \bibfield{journal}{%
  \bibinfo {journal} {J. Chem. Phys.}\ }%
  \textbf{\bibinfo {volume} {124}},\ \bibinfo {pages} {154709} (\bibinfo {year}
  {2006})%
  \bibAnnoteFile{NoStop}{paier-154709}%
\bibitem{bader}%
  \BibitemOpen
  \bibfield{author}{%
  \bibinfo {author} {\bibfnamefont{R.}~\bibnamefont{Bader}},\ }%
  \emph{\bibinfo {title} {Atoms in Molecules: A Quantum Theory}}\ (\bibinfo
  {publisher} {Oxford University Press},\ \bibinfo {address} {New York},\
  \bibinfo {year} {1990})%
  \bibAnnoteFile{NoStop}{bader}%
\bibitem{bader-vasp}%
  \BibitemOpen
  \bibfield{author}{%
  \bibinfo {author} {\bibfnamefont{E.}~\bibnamefont{Sanville}}, \bibinfo
  {author} {\bibfnamefont{S.~D.}\ \bibnamefont{Kenny}}, \bibinfo {author}
  {\bibfnamefont{R.}~\bibnamefont{Smith}},\ and\ \bibinfo {author}
  {\bibfnamefont{G.}~\bibnamefont{Henkelman}},\ }%
  \bibfield{journal}{%
  \bibinfo {journal} {J. Comp. Chem.}\ }%
  \textbf{\bibinfo {volume} {28}},\ \bibinfo {pages} {899} (\bibinfo {year}
  {2007})%
  \bibAnnoteFile{NoStop}{bader-vasp}%
\bibitem{jansen-10026}%
  \BibitemOpen
  \bibfield{author}{%
  \bibinfo {author} {\bibfnamefont{M.}~\bibnamefont{Jansen}}\ and\ \bibinfo
  {author} {\bibfnamefont{U.}~\bibnamefont{Wedig}},\ }%
  \bibfield{journal}{%
  \bibinfo {journal} {Ange. Chemie}\ }%
  \textbf{\bibinfo {volume} {47}} (\bibinfo {year} {2008})%
  \bibAnnoteFile{NoStop}{jansen-10026}%
\bibitem{catlow-4321}%
  \BibitemOpen
  \bibfield{author}{%
  \bibinfo {author} {\bibfnamefont{C.~R.~A.}\ \bibnamefont{Catlow}}\ and\
  \bibinfo {author} {\bibfnamefont{A.}~\bibnamefont{Stoneham}},\ }%
  \bibfield{journal}{%
  \bibinfo {journal} {J. Phys. C}\ }%
  \textbf{\bibinfo {volume} {16}},\ \bibinfo {pages} {4321} (\bibinfo {year}
  {1983})%
  \bibAnnoteFile{NoStop}{catlow-4321}%
\bibitem{ozgur-041301}%
  \BibitemOpen
  \bibfield{author}{%
  \bibinfo {author} {\bibfnamefont{U.}~\bibnamefont{\"{O}zg\"{u}r}}, \bibinfo
  {author} {\bibfnamefont{Y.~I.}\ \bibnamefont{Alivov}}, \bibinfo {author}
  {\bibfnamefont{C.}~\bibnamefont{Liu}}, \bibinfo {author}
  {\bibfnamefont{A.}~\bibnamefont{Teke}}, \bibinfo {author}
  {\bibfnamefont{M.~A.}\ \bibnamefont{Reshchikov}}, \bibinfo {author}
  {\bibfnamefont{S.}~\bibnamefont{Do\u{g}an}}, \bibinfo {author}
  {\bibfnamefont{V.}~\bibnamefont{Avrutin}}, \bibinfo {author}
  {\bibfnamefont{S.-J.}\ \bibnamefont{Cho}},\ and\ \bibinfo {author}
  {\bibfnamefont{H.}~\bibnamefont{Morko\c{c}}},\ }%
  \bibfield{journal}{%
  \Doi{10.1063/1.1992666}{\bibinfo {journal} {J. Appl. Phys.}}\ }%
  \textbf{\bibinfo {volume} {98}},\ \bibinfo {eid} {041301} (\bibinfo {year}
  {2005})%
  \bibAnnoteFile{NoStop}{ozgur-041301}%
\bibitem{kittilstved-291}%
  \BibitemOpen
  \bibfield{author}{%
  \bibinfo {author} {\bibfnamefont{K.~R.}\ \bibnamefont{Kittilstved}}, \bibinfo
  {author} {\bibfnamefont{W.~K.}\ \bibnamefont{Liu}},\ and\ \bibinfo {author}
  {\bibfnamefont{D.~R.}\ \bibnamefont{Gamelin}},\ }%
  \bibfield{journal}{%
  \bibinfo {journal} {Nat. Mater.}\ }%
  \textbf{\bibinfo {volume} {5}},\ \bibinfo {pages} {291} (\bibinfo {year}
  {2006})%
  \bibAnnoteFile{NoStop}{kittilstved-291}%
\bibitem{walsh-256401}%
  \BibitemOpen
  \bibfield{author}{%
  \bibinfo {author} {\bibfnamefont{A.}~\bibnamefont{Walsh}}, \bibinfo {author}
  {\bibfnamefont{J.~L.~F.}\ \bibnamefont{Da~Silva}},\ and\ \bibinfo {author}
  {\bibfnamefont{S.-H.}\ \bibnamefont{Wei}},\ }%
  \bibfield{journal}{%
  \bibinfo {journal} {Phys. Rev. Lett.}\ }%
  \textbf{\bibinfo {volume} {100}},\ \bibinfo {pages} {256401} (\bibinfo {year}
  {2008})%
  \bibAnnoteFile{NoStop}{walsh-256401}%
\bibitem{walsh-159702}%
  \BibitemOpen
  \bibfield{author}{%
  \bibinfo {author} {\bibfnamefont{A.}~\bibnamefont{Walsh}}, \bibinfo {author}
  {\bibfnamefont{J.~L.~F.}\ \bibnamefont{Da~Silva}},\ and\ \bibinfo {author}
  {\bibfnamefont{S.-H.}\ \bibnamefont{Wei}},\ }%
  \bibfield{journal}{%
  \Doi{10.1103/PhysRevLett.102.159702}{\bibinfo {journal} {Phys. Rev. Lett.}}\
  }%
  \textbf{\bibinfo {volume} {102}},\ \bibinfo {pages} {159702} (\bibinfo {year}
  {2009})%
  \bibAnnoteFile{NoStop}{walsh-159702}%
\bibitem{madelung-04}%
  \BibitemOpen
  \bibfield{author}{%
  \bibinfo {author} {\bibfnamefont{O.~M.}\ \bibnamefont{Madelung}},\ }%
  \emph{\bibinfo {title} {Semiconductors: Data Handbook}},\ \bibinfo {edition}
  {3rd}\ ed.\ (\bibinfo {publisher} {Springer},\ \bibinfo {address} {Berlin},\
  \bibinfo {year} {2004})%
  \bibAnnoteFile{NoStop}{madelung-04}%
\bibitem{catlow-1923}%
  \BibitemOpen
  \bibfield{author}{%
  \bibinfo {author} {\bibfnamefont{C.~R.~A.}\ \bibnamefont{Catlow}}, \bibinfo
  {author} {\bibfnamefont{Z.~X.}\ \bibnamefont{Guo}}, \bibinfo {author}
  {\bibfnamefont{M.}~\bibnamefont{Miskufova}}, \bibinfo {author}
  {\bibfnamefont{S.~A.}\ \bibnamefont{Shevlin}}, \bibinfo {author}
  {\bibfnamefont{A.~G.~H.}\ \bibnamefont{Smith}}, \bibinfo {author}
  {\bibfnamefont{A.~A.}\ \bibnamefont{Sokol}}, \bibinfo {author}
  {\bibfnamefont{A.}~\bibnamefont{Walsh}}, \bibinfo {author}
  {\bibfnamefont{D.~J.}\ \bibnamefont{Wilson}},\ and\ \bibinfo {author}
  {\bibfnamefont{S.~M.}\ \bibnamefont{Woodley}},\ }%
  \bibfield{journal}{%
  \bibinfo {journal} {Phil. Trans. Roy. Soc. A}\ }%
  \textbf{\bibinfo {volume} {368}},\ \bibinfo {pages} {1923} (\bibinfo {year}
  {2010})%
  \bibAnnoteFile{NoStop}{catlow-1923}%
\bibitem{sokol-267}%
  \BibitemOpen
  \bibfield{author}{%
  \bibinfo {author} {\bibfnamefont{A.~A.}\ \bibnamefont{Sokol}}, \bibinfo
  {author} {\bibfnamefont{S.~A.}\ \bibnamefont{French}}, \bibinfo {author}
  {\bibfnamefont{S.~T.}\ \bibnamefont{Bromley}}, \bibinfo {author}
  {\bibfnamefont{C.~R.~A.}\ \bibnamefont{Catlow}}, \bibinfo {author}
  {\bibfnamefont{H.~J.~J.}\ \bibnamefont{van Dam}},\ and\ \bibinfo {author}
  {\bibfnamefont{P.}~\bibnamefont{Sherwood}},\ }%
  \bibfield{journal}{%
  \bibinfo {journal} {Faraday Discuss.}\ }%
  \textbf{\bibinfo {volume} {134}},\ \bibinfo {pages} {267} (\bibinfo {year}
  {2007})%
  \bibAnnoteFile{NoStop}{sokol-267}%
\bibitem{hamberg-r123}%
  \BibitemOpen
  \bibfield{author}{%
  \bibinfo {author} {\bibfnamefont{I.}~\bibnamefont{Hamberg}}\ and\ \bibinfo
  {author} {\bibfnamefont{C.~G.}\ \bibnamefont{Granqvist}},\ }%
  \bibfield{journal}{%
  \bibinfo {journal} {J. Appl. Phys.}\ }%
  \textbf{\bibinfo {volume} {60}},\ \bibinfo {pages} {R123} (\bibinfo {year}
  {1986})%
  \bibAnnoteFile{NoStop}{hamberg-r123}%
\bibitem{marezio-723}%
  \BibitemOpen
  \bibfield{author}{%
  \bibinfo {author} {\bibfnamefont{M.}~\bibnamefont{Marezio}},\ }%
  \bibfield{journal}{%
  \bibinfo {journal} {Acta Crystallogr.}\ }%
  \textbf{\bibinfo {volume} {20}},\ \bibinfo {pages} {723} (\bibinfo {year}
  {1966})%
  \bibAnnoteFile{NoStop}{marezio-723}%
\bibitem{fuchs-155107}%
  \BibitemOpen
  \bibfield{author}{%
  \bibinfo {author} {\bibfnamefont{F.}~\bibnamefont{Fuchs}}\ and\ \bibinfo
  {author} {\bibfnamefont{F.}~\bibnamefont{Bechstedt}},\ }%
  \bibfield{journal}{%
  \bibinfo {journal} {Phys. Rev. B}\ }%
  \textbf{\bibinfo {volume} {77}},\ \bibinfo {pages} {155107} (\bibinfo {year}
  {2008})%
  \bibAnnoteFile{NoStop}{fuchs-155107}%
\bibitem{passlack-686}%
  \BibitemOpen
  \bibfield{author}{%
  \bibinfo {author} {\bibfnamefont{M.}~\bibnamefont{Passlack}}, \bibinfo
  {author} {\bibfnamefont{E.~F.}\ \bibnamefont{Schubert}}, \bibinfo {author}
  {\bibfnamefont{W.~S.}\ \bibnamefont{Hobson}}, \bibinfo {author}
  {\bibfnamefont{M.}~\bibnamefont{Hong}}, \bibinfo {author}
  {\bibfnamefont{N.}~\bibnamefont{Moriya}}, \bibinfo {author}
  {\bibfnamefont{S.~N.~G.}\ \bibnamefont{Chu}}, \bibinfo {author}
  {\bibfnamefont{K.}~\bibnamefont{Konstadinidis}}, \bibinfo {author}
  {\bibfnamefont{J.~P.}\ \bibnamefont{Mannaerts}}, \bibinfo {author}
  {\bibfnamefont{M.~L.}\ \bibnamefont{Schnoes}},\ and\ \bibinfo {author}
  {\bibfnamefont{G.~J.}\ \bibnamefont{Zydzik}},\ }%
  \bibfield{journal}{%
  \Doi{10.1063/1.359055}{\bibinfo {journal} {J. Appl. Phys.}}\ }%
  \textbf{\bibinfo {volume} {77}},\ \bibinfo {pages} {686} (\bibinfo {year}
  {1995})%
  \bibAnnoteFile{NoStop}{passlack-686}%
\bibitem{geller-676}%
  \BibitemOpen
  \bibfield{author}{%
  \bibinfo {author} {\bibfnamefont{S.}~\bibnamefont{Geller}},\ }%
  \bibfield{journal}{%
  \Doi{10.1063/1.1731237}{\bibinfo {journal} {J. Chem. Phys.}}\ }%
  \textbf{\bibinfo {volume} {33}},\ \bibinfo {pages} {676} (\bibinfo {year}
  {1960})%
  \bibAnnoteFile{NoStop}{geller-676}%
\bibitem{varley-142106}%
  \BibitemOpen
  \bibfield{author}{%
  \bibinfo {author} {\bibfnamefont{J.~B.}\ \bibnamefont{Varley}}, \bibinfo
  {author} {\bibfnamefont{J.~R.}\ \bibnamefont{Weber}}, \bibinfo {author}
  {\bibfnamefont{A.}~\bibnamefont{Janotti}},\ and\ \bibinfo {author}
  {\bibfnamefont{C.~G.~V.}\ \bibnamefont{de~Walle}},\ }%
  \bibfield{journal}{%
  \Doi{10.1063/1.3499306}{\bibinfo {journal} {Appl. Phys. Lett.}}\ }%
  \textbf{\bibinfo {volume} {97}},\ \bibinfo {eid} {142106} (\bibinfo {year}
  {2010})%
  \bibAnnoteFile{NoStop}{varley-142106}%
\bibitem{evans-1995}%
  \BibitemOpen
  \bibfield{author}{%
  \bibinfo {author} {\bibfnamefont{B.}~\bibnamefont{Evans}},\ }%
  \bibfield{journal}{%
  \bibinfo {journal} {J. Nucl. Mater.}\ }%
  \textbf{\bibinfo {volume} {219}},\ \bibinfo {pages} {202} (\bibinfo {year}
  {1995})%
  \bibAnnoteFile{NoStop}{evans-1995}%
\bibitem{tomiki-573}%
  \BibitemOpen
  \bibfield{author}{%
  \bibinfo {author} {\bibfnamefont{T.}~\bibnamefont{Tomiki}}, \bibinfo {author}
  {\bibfnamefont{Y.}~\bibnamefont{Ganaha}}, \bibinfo {author}
  {\bibfnamefont{T.}~\bibnamefont{Shikenbaru}}, \bibinfo {author}
  {\bibfnamefont{T.}~\bibnamefont{Futemma}}, \bibinfo {author}
  {\bibfnamefont{M.}~\bibnamefont{Yuri}}, \bibinfo {author}
  {\bibfnamefont{Y.}~\bibnamefont{Aiura}}, \bibinfo {author}
  {\bibfnamefont{S.}~\bibnamefont{Sato}}, \bibinfo {author}
  {\bibfnamefont{H.}~\bibnamefont{Fukutani}}, \bibinfo {author}
  {\bibfnamefont{H.}~\bibnamefont{Kato}}, \bibinfo {author}
  {\bibfnamefont{T.}~\bibnamefont{Miyahara}}, \bibinfo {author}
  {\bibfnamefont{A.}~\bibnamefont{Yonesu}},\ and\ \bibinfo {author}
  {\bibfnamefont{J.}~\bibnamefont{Tamashiro}},\ }%
  \bibfield{journal}{%
  \Doi{10.1143/JPSJ.62.573}{\bibinfo {journal} {J. Phys. Soc. Jpn.}}\ }%
  \textbf{\bibinfo {volume} {62}},\ \bibinfo {pages} {573} (\bibinfo {year}
  {1993})%
  \bibAnnoteFile{NoStop}{tomiki-573}%
\bibitem{thompson-79}%
  \BibitemOpen
  \bibfield{author}{%
  \bibinfo {author} {\bibfnamefont{P.}~\bibnamefont{Thompson}}, \bibinfo
  {author} {\bibfnamefont{D.}~\bibnamefont{Cox}},\ and\ \bibinfo {author}
  {\bibfnamefont{J.}~\bibnamefont{Hastings}},\ }%
  \bibfield{journal}{%
  \bibinfo {journal} {J. Appl. Crystal.}\ }%
  \textbf{\bibinfo {volume} {20}},\ \bibinfo {pages} {79} (\bibinfo {year}
  {1987})%
  \bibAnnoteFile{NoStop}{thompson-79}%
\bibitem{weber-1756}%
  \BibitemOpen
  \bibfield{author}{%
  \bibinfo {author} {\bibfnamefont{J.}~\bibnamefont{Weber}}, \bibinfo {author}
  {\bibfnamefont{A.}~\bibnamefont{Janotti}},\ and\ \bibinfo {author}
  {\bibfnamefont{C.~V.}\ \bibnamefont{de~Walle}},\ }%
  \bibfield{journal}{%
  \Doi{DOI: 10.1016/j.mee.2009.03.059}{\bibinfo {journal} {Microelect. Eng.}}\
  }%
  \textbf{\bibinfo {volume} {86}},\ \bibinfo {pages} {1756} (\bibinfo {year}
  {2009})%
  \bibAnnoteFile{NoStop}{weber-1756}%
\bibitem{catlow-1006}%
  \BibitemOpen
  \bibfield{author}{%
  \bibinfo {author} {\bibfnamefont{C.~R.~A.}\ \bibnamefont{Catlow}}, \bibinfo
  {author} {\bibfnamefont{R.}~\bibnamefont{James}}, \bibinfo {author}
  {\bibfnamefont{W.~C.}\ \bibnamefont{Mackrodt}},\ and\ \bibinfo {author}
  {\bibfnamefont{R.~F.}\ \bibnamefont{Stewart}},\ }%
  \bibfield{journal}{%
  \Doi{10.1103/PhysRevB.25.1006}{\bibinfo {journal} {Phys. Rev. B}}\ }%
  \textbf{\bibinfo {volume} {25}},\ \bibinfo {pages} {1006} (\bibinfo {year}
  {1982})%
  \bibAnnoteFile{NoStop}{catlow-1006}%
\bibitem{hine-114111}%
  \BibitemOpen
  \bibfield{author}{%
  \bibinfo {author} {\bibfnamefont{N.}~\bibnamefont{Hine}}, \bibinfo {author}
  {\bibfnamefont{P.}~\bibnamefont{Haynes}}, \bibinfo {author}
  {\bibfnamefont{A.}~\bibnamefont{Mostofi}},\ and\ \bibinfo {author}
  {\bibfnamefont{M.}~\bibnamefont{Payne}},\ }%
  \bibfield{journal}{%
  \bibinfo {journal} {J. Chem. Phys.}\ }%
  \textbf{\bibinfo {volume} {133}},\ \bibinfo {pages} {114111} (\bibinfo {year}
  {2010})%
  \bibAnnoteFile{NoStop}{hine-114111}%
\bibitem{sokol-44}%
  \BibitemOpen
  \bibfield{author}{%
  \bibinfo {author} {\bibfnamefont{A.~A.}\ \bibnamefont{Sokol}}, \bibinfo
  {author} {\bibfnamefont{A.}~\bibnamefont{Walsh}},\ and\ \bibinfo {author}
  {\bibfnamefont{C.~R.~A.}\ \bibnamefont{Catlow}},\ }%
  \bibfield{journal}{%
  \Doi{DOI: 10.1016/j.cplett.2010.04.029}{\bibinfo {journal} {Chem. Phys.
  Lett.}}\ }%
  \textbf{\bibinfo {volume} {492}},\ \bibinfo {pages} {44 } (\bibinfo {year}
  {2010})%
  \bibAnnoteFile{NoStop}{sokol-44}%
\bibitem{crc}%
  \BibitemOpen
  \bibfield{author}{%
  \bibinfo {author} {\bibfnamefont{D.~R.}\ \bibnamefont{Lide}},\ }%
  \emph{\bibinfo {title} {CRC Handbook}},\ \bibinfo {edition} {83rd}\ ed.\
  (\bibinfo {publisher} {CRC Press},\ \bibinfo {address} {London},\ \bibinfo
  {year} {2002})%
  \bibAnnoteFile{NoStop}{crc}%
\bibitem{orita-4166}%
  \BibitemOpen
  \bibfield{author}{%
  \bibinfo {author} {\bibfnamefont{M.}~\bibnamefont{Orita}}, \bibinfo {author}
  {\bibfnamefont{H.}~\bibnamefont{Ohta}}, \bibinfo {author}
  {\bibfnamefont{M.}~\bibnamefont{Hirano}},\ and\ \bibinfo {author}
  {\bibfnamefont{H.}~\bibnamefont{Hosono}},\ }%
  \bibfield{journal}{%
  \Doi{10.1063/1.1330559}{\bibinfo {journal} {Appl. Phys. Lett.}}\ }%
  \textbf{\bibinfo {volume} {77}},\ \bibinfo {pages} {4166} (\bibinfo {year}
  {2000})%
  \bibAnnoteFile{NoStop}{orita-4166}%
\bibitem{Kimizuka-1995-170}%
  \BibitemOpen
  \bibfield{author}{%
  \bibinfo {author} {\bibfnamefont{N.}~\bibnamefont{Kimizuka}}, \bibinfo
  {author} {\bibfnamefont{M.}~\bibnamefont{Isobe}},\ and\ \bibinfo {author}
  {\bibfnamefont{M.}~\bibnamefont{Nakamura}},\ }%
  \bibfield{journal}{%
  \bibinfo {journal} {J.\ Solid State Chem.}\ }%
  \textbf{\bibinfo {volume} {116}},\ \bibinfo {pages} {170} (\bibinfo {year}
  {1995})%
  \bibAnnoteFile{NoStop}{Kimizuka-1995-170}%
\bibitem{Phani-1998-3969}%
  \BibitemOpen
  \bibfield{author}{%
  \bibinfo {author} {\bibfnamefont{A.~R.}\ \bibnamefont{Phani}}, \bibinfo
  {author} {\bibfnamefont{S.}~\bibnamefont{Santucci}}, \bibinfo {author}
  {\bibfnamefont{S.}~\bibnamefont{{Di~Nardo}}}, \bibinfo {author}
  {\bibfnamefont{L.}~\bibnamefont{Lozzi}}, \bibinfo {author}
  {\bibfnamefont{M.}~\bibnamefont{Passacantando}}, \bibinfo {author}
  {\bibfnamefont{P.}~\bibnamefont{Picozzi}},\ and\ \bibinfo {author}
  {\bibfnamefont{C.}~\bibnamefont{Cantalini}},\ }%
  \bibfield{journal}{%
  \bibinfo {journal} {J.\ Mater.\ Sci.}\ }%
  \textbf{\bibinfo {volume} {33}},\ \bibinfo {pages} {3969} (\bibinfo {year}
  {1998})%
  \bibAnnoteFile{NoStop}{Phani-1998-3969}%
\bibitem{Li-1999-355}%
  \BibitemOpen
  \bibfield{author}{%
  \bibinfo {author} {\bibfnamefont{C.}~\bibnamefont{Li}}, \bibinfo {author}
  {\bibfnamefont{Y.}~\bibnamefont{Bando}}, \bibinfo {author}
  {\bibfnamefont{M.}~\bibnamefont{Nakamura}}, \bibinfo {author}
  {\bibfnamefont{K.}~\bibnamefont{Kurashima}},\ and\ \bibinfo {author}
  {\bibfnamefont{N.}~\bibnamefont{Kimizuka}},\ }%
  \bibfield{journal}{%
  \bibinfo {journal} {Acta.\ Cryst.}\ }%
  \textbf{\bibinfo {volume} {B55}},\ \bibinfo {pages} {355} (\bibinfo {year}
  {1999})%
  \bibAnnoteFile{NoStop}{Li-1999-355}%
\bibitem{Kim-2004-163}%
  \BibitemOpen
  \bibfield{author}{%
  \bibinfo {author} {\bibfnamefont{J.~S.}\ \bibnamefont{Kim}}, \bibinfo
  {author} {\bibfnamefont{H.~L.}\ \bibnamefont{Park}}, \bibinfo {author}
  {\bibfnamefont{C.~M.}\ \bibnamefont{Chon}}, \bibinfo {author}
  {\bibfnamefont{H.~S.}\ \bibnamefont{Moon}},\ and\ \bibinfo {author}
  {\bibfnamefont{T.~W.}\ \bibnamefont{Kim}},\ }%
  \bibfield{journal}{%
  \bibinfo {journal} {Solid State Commun.}\ }%
  \textbf{\bibinfo {volume} {129}},\ \bibinfo {pages} {163} (\bibinfo {year}
  {2004})%
  \bibAnnoteFile{NoStop}{Kim-2004-163}%
\bibitem{Michiue-2008-521}%
  \BibitemOpen
  \bibfield{author}{%
  \bibinfo {author} {\bibfnamefont{Y.}~\bibnamefont{Michiue}}, \bibinfo
  {author} {\bibfnamefont{N.}~\bibnamefont{Kimizuka}},\ and\ \bibinfo {author}
  {\bibfnamefont{Y.}~\bibnamefont{Kanke}},\ }%
  \bibfield{journal}{%
  \bibinfo {journal} {Acta Cryst.}\ }%
  \textbf{\bibinfo {volume} {B64}},\ \bibinfo {pages} {521} (\bibinfo {year}
  {2008})%
  \bibAnnoteFile{NoStop}{Michiue-2008-521}%
\bibitem{yoshioka-014309}%
  \BibitemOpen
  \bibfield{author}{%
  \bibinfo {author} {\bibfnamefont{S.}~\bibnamefont{Yoshioka}}, \bibinfo
  {author} {\bibfnamefont{F.}~\bibnamefont{Oba}}, \bibinfo {author}
  {\bibfnamefont{R.}~\bibnamefont{Huang}}, \bibinfo {author}
  {\bibfnamefont{I.}~\bibnamefont{Tanaka}}, \bibinfo {author}
  {\bibfnamefont{T.}~\bibnamefont{Mizoguchi}},\ and\ \bibinfo {author}
  {\bibfnamefont{T.}~\bibnamefont{Yamamoto}},\ }%
  \bibfield{journal}{%
  \bibinfo {journal} {J. Appl. Phys}\ }%
  \textbf{\bibinfo {volume} {103}},\ \bibinfo {pages} {014309} (\bibinfo {year}
  {2008})%
  \bibAnnoteFile{NoStop}{yoshioka-014309}%
\bibitem{yoshioka-137}%
  \BibitemOpen
  \bibfield{author}{%
  \bibinfo {author} {\bibfnamefont{S.}~\bibnamefont{Yoshioka}}, \bibinfo
  {author} {\bibfnamefont{K.}~\bibnamefont{Toyoura}}, \bibinfo {author}
  {\bibfnamefont{F.}~\bibnamefont{Oba}}, \bibinfo {author}
  {\bibfnamefont{A.}~\bibnamefont{Kuwabara}}, \bibinfo {author}
  {\bibfnamefont{K.}~\bibnamefont{Matsunaga}},\ and\ \bibinfo {author}
  {\bibfnamefont{I.}~\bibnamefont{Tanaka}},\ }%
  \bibfield{journal}{%
  \bibinfo {journal} {Journal of Solid State Chemistry}\ }%
  \textbf{\bibinfo {volume} {181}},\ \bibinfo {pages} {137} (\bibinfo {year}
  {2008})%
  \bibAnnoteFile{NoStop}{yoshioka-137}%
\bibitem{vinnichenko-141907}%
  \BibitemOpen
  \bibfield{author}{%
  \bibinfo {author} {\bibfnamefont{M.}~\bibnamefont{Vinnichenko}}, \bibinfo
  {author} {\bibfnamefont{R.}~\bibnamefont{Gago}}, \bibinfo {author}
  {\bibfnamefont{S.}~\bibnamefont{Cornelius}}, \bibinfo {author}
  {\bibfnamefont{N.}~\bibnamefont{Shevchenko}}, \bibinfo {author}
  {\bibfnamefont{A.}~\bibnamefont{Rogozin}}, \bibinfo {author}
  {\bibfnamefont{A.}~\bibnamefont{Kolitsch}}, \bibinfo {author}
  {\bibfnamefont{F.}~\bibnamefont{Munnik}},\ and\ \bibinfo {author}
  {\bibfnamefont{W.}~\bibnamefont{Moller}},\ }%
  \bibfield{journal}{%
  \bibinfo {journal} {Appl. Phys. Lett.}\ }%
  \textbf{\bibinfo {volume} {96}},\ \bibinfo {pages} {141907} (\bibinfo {year}
  {2010})%
  \bibAnnoteFile{NoStop}{vinnichenko-141907}%
\bibitem{horwat-132003}%
  \BibitemOpen
  \bibfield{author}{%
  \bibinfo {author} {\bibfnamefont{D.}~\bibnamefont{Horwat}}, \bibinfo {author}
  {\bibfnamefont{M.}~\bibnamefont{Jullien}}, \bibinfo {author}
  {\bibfnamefont{F.}~\bibnamefont{Capon}}, \bibinfo {author}
  {\bibfnamefont{J.}~\bibnamefont{Pierson}}, \bibinfo {author}
  {\bibfnamefont{J.}~\bibnamefont{Andersson}},\ and\ \bibinfo {author}
  {\bibfnamefont{J.}~\bibnamefont{Endrino}},\ }%
  \bibfield{journal}{%
  \bibinfo {journal} {J. Phys. D: Appl. Phys.}\ }%
  \textbf{\bibinfo {volume} {43}},\ \bibinfo {pages} {132003} (\bibinfo {year}
  {2010})%
  \bibAnnoteFile{NoStop}{horwat-132003}%
\bibitem{vesta}%
  \BibitemOpen
  \bibfield{author}{%
  \bibinfo {author} {\bibfnamefont{K.}~\bibnamefont{Momma}}\ and\ \bibinfo
  {author} {\bibfnamefont{F.}~\bibnamefont{Izumi}},\ }%
  \bibfield{journal}{%
  \bibinfo {journal} {J. Appl. Cryst.}\ }%
  \textbf{\bibinfo {volume} {41}},\ \bibinfo {pages} {653} (\bibinfo {year}
  {2008})%
  \bibAnnoteFile{NoStop}{vesta}%
\bibitem{Kasper-1967-113}%
  \BibitemOpen
  \bibfield{author}{%
  \bibinfo {author} {\bibfnamefont{H.}~\bibnamefont{Kasper}},\ }%
  \bibfield{journal}{%
  \bibinfo {journal} {Z.\ Anorg.\ Allg.\ Chem.}\ }%
  \textbf{\bibinfo {volume} {349}},\ \bibinfo {pages} {113} (\bibinfo {year}
  {1967})%
  \bibAnnoteFile{NoStop}{Kasper-1967-113}%
\bibitem{Cannard-1988-418}%
  \BibitemOpen
  \bibfield{author}{%
  \bibinfo {author} {\bibfnamefont{P.~J.}\ \bibnamefont{Cannard}}\ and\
  \bibinfo {author} {\bibfnamefont{R.~J.~D.}\ \bibnamefont{Tilley}},\ }%
  \bibfield{journal}{%
  \bibinfo {journal} {J.\ Solid State Chem.}\ }%
  \textbf{\bibinfo {volume} {73}},\ \bibinfo {pages} {418} (\bibinfo {year}
  {1988})%
  \bibAnnoteFile{NoStop}{Cannard-1988-418}%
\bibitem{Kimizuka-1981-109}%
  \BibitemOpen
  \bibfield{author}{%
  \bibinfo {author} {\bibfnamefont{N.}~\bibnamefont{Kimizuka}}\ and\ \bibinfo
  {author} {\bibfnamefont{E.}~\bibnamefont{Takayama}},\ }%
  \bibfield{journal}{%
  \bibinfo {journal} {J.\ Solid State Chem.}\ }%
  \textbf{\bibinfo {volume} {40}},\ \bibinfo {pages} {109} (\bibinfo {year}
  {1981})%
  \bibAnnoteFile{NoStop}{Kimizuka-1981-109}%
\bibitem{Isobe-1994-332}%
  \BibitemOpen
  \bibfield{author}{%
  \bibinfo {author} {\bibfnamefont{M.}~\bibnamefont{Isobe}}, \bibinfo {author}
  {\bibfnamefont{N.}~\bibnamefont{Kimizuka}}, \bibinfo {author}
  {\bibfnamefont{M.}~\bibnamefont{Nakamura}},\ and\ \bibinfo {author}
  {\bibfnamefont{T.}~\bibnamefont{Mohri}},\ }%
  \bibfield{journal}{%
  \bibinfo {journal} {Acta.\ Cryst.}\ }%
  \textbf{\bibinfo {volume} {C50}},\ \bibinfo {pages} {332} (\bibinfo {year}
  {1994})%
  \bibAnnoteFile{NoStop}{Isobe-1994-332}%
\bibitem{Yan-1998-2585}%
  \BibitemOpen
  \bibfield{author}{%
  \bibinfo {author} {\bibfnamefont{Y.}~\bibnamefont{Yan}}, \bibinfo {author}
  {\bibfnamefont{S.~J.}\ \bibnamefont{Pennycook}}, \bibinfo {author}
  {\bibfnamefont{J.}~\bibnamefont{Dai}}, \bibinfo {author}
  {\bibfnamefont{R.~P.~H.}\ \bibnamefont{Chang}}, \bibinfo {author}
  {\bibfnamefont{A.}~\bibnamefont{Wang}},\ and\ \bibinfo {author}
  {\bibfnamefont{T.~J.}\ \bibnamefont{Marks}},\ }%
  \bibfield{journal}{%
  \bibinfo {journal} {Appl.\ Phys.\ Lett.}\ }%
  \textbf{\bibinfo {volume} {73}},\ \bibinfo {pages} {2585} (\bibinfo {year}
  {1998})%
  \bibAnnoteFile{NoStop}{Yan-1998-2585}%
\bibitem{Moriga-1998-1310}%
  \BibitemOpen
  \bibfield{author}{%
  \bibinfo {author} {\bibfnamefont{T.}~\bibnamefont{Moriga}}, \bibinfo {author}
  {\bibfnamefont{D.~D.}\ \bibnamefont{Edwards}}, \bibinfo {author}
  {\bibfnamefont{T.~O.}\ \bibnamefont{Mason}}, \bibinfo {author}
  {\bibfnamefont{G.~B.}\ \bibnamefont{Palmer}}, \bibinfo {author}
  {\bibfnamefont{K.~R.}\ \bibnamefont{Poeppelmeier}}, \bibinfo {author}
  {\bibfnamefont{J.~L.}\ \bibnamefont{Schindler}}, \bibinfo {author}
  {\bibfnamefont{C.~R.}\ \bibnamefont{Kannewurf}},\ and\ \bibinfo {author}
  {\bibfnamefont{I.}~\bibnamefont{Nakabayashi}},\ }%
  \bibfield{journal}{%
  \bibinfo {journal} {J.\ Am.\ Ceram.\ Soc.}\ }%
  \textbf{\bibinfo {volume} {81}},\ \bibinfo {pages} {1310} (\bibinfo {year}
  {1998})%
  \bibAnnoteFile{NoStop}{Moriga-1998-1310}%
\bibitem{Uchida-1994-146}%
  \BibitemOpen
  \bibfield{author}{%
  \bibinfo {author} {\bibfnamefont{N.}~\bibnamefont{Uchida}}, \bibinfo {author}
  {\bibfnamefont{Y.}~\bibnamefont{Bando}}, \bibinfo {author}
  {\bibfnamefont{M.}~\bibnamefont{Nakamura}},\ and\ \bibinfo {author}
  {\bibfnamefont{N.}~\bibnamefont{Kimizuka}},\ }%
  \bibfield{journal}{%
  \bibinfo {journal} {J.\ Electron.\ Microsc.}\ }%
  \textbf{\bibinfo {volume} {43}},\ \bibinfo {pages} {146} (\bibinfo {year}
  {1994})%
  \bibAnnoteFile{NoStop}{Uchida-1994-146}%
\bibitem{Li-1998-347}%
  \BibitemOpen
  \bibfield{author}{%
  \bibinfo {author} {\bibfnamefont{C.}~\bibnamefont{Li}}, \bibinfo {author}
  {\bibfnamefont{Y.}~\bibnamefont{Bando}}, \bibinfo {author}
  {\bibfnamefont{M.}~\bibnamefont{Nakamura}}, \bibinfo {author}
  {\bibfnamefont{M.}~\bibnamefont{Onoda}},\ and\ \bibinfo {author}
  {\bibfnamefont{N.}~\bibnamefont{Kimizuka}},\ }%
  \bibfield{journal}{%
  \bibinfo {journal} {J.\ Solid State Chem.}\ }%
  \textbf{\bibinfo {volume} {139}},\ \bibinfo {pages} {347} (\bibinfo {year}
  {1998})%
  \bibAnnoteFile{NoStop}{Li-1998-347}%
\bibitem{Li-2000-543}%
  \BibitemOpen
  \bibfield{author}{%
  \bibinfo {author} {\bibfnamefont{C.}~\bibnamefont{Li}}, \bibinfo {author}
  {\bibfnamefont{Y.}~\bibnamefont{Bando}}, \bibinfo {author}
  {\bibfnamefont{M.}~\bibnamefont{Nakamura}},\ and\ \bibinfo {author}
  {\bibfnamefont{N.}~\bibnamefont{Kimizuka}},\ }%
  \bibfield{journal}{%
  \bibinfo {journal} {Micron}\ }%
  \textbf{\bibinfo {volume} {31}},\ \bibinfo {pages} {543} (\bibinfo {year}
  {2000})%
  \bibAnnoteFile{NoStop}{Li-2000-543}%
\bibitem{Wolf-2007-549}%
  \BibitemOpen
  \bibfield{author}{%
  \bibinfo {author} {\bibfnamefont{F.}~\bibnamefont{Wolf}}, \bibinfo {author}
  {\bibfnamefont{B.~H.}\ \bibnamefont{Freitag}},\ and\ \bibinfo {author}
  {\bibfnamefont{W.}~\bibnamefont{Mader}},\ }%
  \bibfield{journal}{%
  \bibinfo {journal} {Micron}\ }%
  \textbf{\bibinfo {volume} {38}},\ \bibinfo {pages} {549} (\bibinfo {year}
  {2007})%
  \bibAnnoteFile{NoStop}{Wolf-2007-549}%
\bibitem{dasilva-255501}%
  \BibitemOpen
  \bibfield{author}{%
  \bibinfo {author} {\bibfnamefont{J.~L. F.~D.}\ \bibnamefont{Silva}}, \bibinfo
  {author} {\bibfnamefont{Y.}~\bibnamefont{Yan}},\ and\ \bibinfo {author}
  {\bibfnamefont{S.-H.}\ \bibnamefont{Wei}},\ }%
  \bibfield{journal}{%
  \Doi{10.1103/PhysRevLett.100.255501}{\bibinfo {journal} {Phys. Rev. Lett.}}\
  }%
  \textbf{\bibinfo {volume} {100}},\ \bibinfo {pages} {255501} (\bibinfo {year}
  {2008})%
  \bibAnnoteFile{NoStop}{dasilva-255501}%
\bibitem{DaSilva-2009-214118}%
  \BibitemOpen
  \bibfield{author}{%
  \bibinfo {author} {\bibfnamefont{J.~L.~F.}\ \bibnamefont{{Da~Silva}}},
  \bibinfo {author} {\bibfnamefont{A.}~\bibnamefont{Walsh}},\ and\ \bibinfo
  {author} {\bibfnamefont{S.-H.}\ \bibnamefont{Wei}},\ }%
  \bibfield{journal}{%
  \bibinfo {journal} {Phys.\ Rev.\ B}\ }%
  \textbf{\bibinfo {volume} {80}},\ \bibinfo {pages} {214118} (\bibinfo {year}
  {2009})%
  \bibAnnoteFile{NoStop}{DaSilva-2009-214118}%
\bibitem{Marezio-1966-723}%
  \BibitemOpen
  \bibfield{author}{%
  \bibinfo {author} {\bibfnamefont{M.}~\bibnamefont{Marezio}},\ }%
  \bibfield{journal}{%
  \bibinfo {journal} {Acta Cryst.}\ }%
  \textbf{\bibinfo {volume} {20}},\ \bibinfo {pages} {723} (\bibinfo {year}
  {1966})%
  \bibAnnoteFile{NoStop}{Marezio-1966-723}%
\bibitem{Giaquinta-1994-5}%
  \BibitemOpen
  \bibfield{author}{%
  \bibinfo {author} {\bibfnamefont{D.~M.}\ \bibnamefont{Giaquinta}}, \bibinfo
  {author} {\bibfnamefont{W.~M.}\ \bibnamefont{Davis}},\ and\ \bibinfo {author}
  {\bibfnamefont{H.-C.}\ \bibnamefont{{zur Loye}}},\ }%
  \bibfield{journal}{%
  \bibinfo {journal} {Acta Cryst.\ C}\ }%
  \textbf{\bibinfo {volume} {50}},\ \bibinfo {pages} {5} (\bibinfo {year}
  {1994})%
  \bibAnnoteFile{NoStop}{Giaquinta-1994-5}%
\bibitem{shannon-751}%
  \BibitemOpen
  \bibfield{author}{%
  \bibinfo {author} {\bibfnamefont{R.~D.}\ \bibnamefont{Shannon}},\ }%
  \bibfield{journal}{%
  \bibinfo {journal} {Acta Crystallogr., Sect. A}\ }%
  \textbf{\bibinfo {volume} {32}},\ \bibinfo {pages} {751} (\bibinfo {year}
  {1976})%
  \bibAnnoteFile{NoStop}{shannon-751}%
\bibitem{hosono-165}%
  \BibitemOpen
  \bibfield{author}{%
  \bibinfo {author} {\bibfnamefont{H.}~\bibnamefont{Hosono}}, \bibinfo {author}
  {\bibfnamefont{N.}~\bibnamefont{Kikuchi}}, \bibinfo {author}
  {\bibfnamefont{N.}~\bibnamefont{Ueda}},\ and\ \bibinfo {author}
  {\bibfnamefont{H.}~\bibnamefont{Kawazoe}},\ }%
  \bibfield{journal}{%
  \bibinfo {journal} {J. Non-Cryst. Solids}\ }%
  \textbf{\bibinfo {volume} {198-200}},\ \bibinfo {pages} {165} (\bibinfo
  {year} {1996})%
  \bibAnnoteFile{NoStop}{hosono-165}%
\bibitem{nomura-1269}%
  \BibitemOpen
  \bibfield{author}{%
  \bibinfo {author} {\bibfnamefont{K.}~\bibnamefont{Nomura}}, \bibinfo {author}
  {\bibfnamefont{H.}~\bibnamefont{Ohta}}, \bibinfo {author}
  {\bibfnamefont{K.}~\bibnamefont{Ueda}}, \bibinfo {author}
  {\bibfnamefont{T.}~\bibnamefont{Kamiya}}, \bibinfo {author}
  {\bibfnamefont{M.}~\bibnamefont{Hirano}},\ and\ \bibinfo {author}
  {\bibfnamefont{H.}~\bibnamefont{Hosono}},\ }%
  \bibfield{journal}{%
  \bibinfo {journal} {Science}\ }%
  \textbf{\bibinfo {volume} {300}},\ \bibinfo {pages} {1269} (\bibinfo {year}
  {2003})%
  \bibAnnoteFile{NoStop}{nomura-1269}%
\bibitem{nomura-202117}%
  \BibitemOpen
  \bibfield{author}{%
  \bibinfo {author} {\bibfnamefont{K.}~\bibnamefont{Nomura}}, \bibinfo {author}
  {\bibfnamefont{T.}~\bibnamefont{Kamiya}}, \bibinfo {author}
  {\bibfnamefont{H.}~\bibnamefont{Yanagi}}, \bibinfo {author}
  {\bibfnamefont{E.}~\bibnamefont{Ikenaga}}, \bibinfo {author}
  {\bibfnamefont{K.}~\bibnamefont{Yang}}, \bibinfo {author}
  {\bibfnamefont{K.}~\bibnamefont{Kobayashi}}, \bibinfo {author}
  {\bibfnamefont{M.}~\bibnamefont{Hirano}},\ and\ \bibinfo {author}
  {\bibfnamefont{H.}~\bibnamefont{Hosono}},\ }%
  \bibfield{journal}{%
  \bibinfo {journal} {Appl. Phys. Lett.}\ }%
  \textbf{\bibinfo {volume} {92}},\ \bibinfo {pages} {202117} (\bibinfo {year}
  {2008})%
  \bibAnnoteFile{NoStop}{nomura-202117}%
\bibitem{ecn-1}%
  \BibitemOpen
  \bibfield{author}{%
  \bibinfo {author} {\bibfnamefont{R.}~\bibnamefont{Hoppe}}, \bibinfo {author}
  {\bibfnamefont{S.}~\bibnamefont{Voigt}}, \bibinfo {author}
  {\bibfnamefont{H.}~\bibnamefont{Glaum}}, \bibinfo {author}
  {\bibfnamefont{J.}~\bibnamefont{Kissel}}, \bibinfo {author}
  {\bibfnamefont{H.~P.}\ \bibnamefont{Muller}},\ and\ \bibinfo {author}
  {\bibfnamefont{K.}~\bibnamefont{Bernet}},\ }%
  \bibfield{journal}{%
  \bibinfo {journal} {J. Less-Common Met.}\ }%
  \textbf{\bibinfo {volume} {156}},\ \bibinfo {pages} {105} (\bibinfo {year}
  {1989})%
  \bibAnnoteFile{NoStop}{ecn-1}%
\bibitem{ecn-2}%
  \BibitemOpen
  \bibfield{author}{%
  \bibinfo {author} {\bibfnamefont{R.}~\bibnamefont{Hoppe}},\ }%
  \bibfield{journal}{%
  \bibinfo {journal} {Angew. Chem. Internat. Edit.}\ }%
  \textbf{\bibinfo {volume} {9}},\ \bibinfo {pages} {25} (\bibinfo {year}
  {1970})%
  \bibAnnoteFile{NoStop}{ecn-2}%
\bibitem{wooten-1392}%
  \BibitemOpen
  \bibfield{author}{%
  \bibinfo {author} {\bibfnamefont{F.}~\bibnamefont{Wooten}}, \bibinfo {author}
  {\bibfnamefont{K.}~\bibnamefont{Winer}},\ and\ \bibinfo {author}
  {\bibfnamefont{D.}~\bibnamefont{Weaire}},\ }%
  \bibfield{journal}{%
  \bibinfo {journal} {Phys. Rev. Lett.}\ }%
  \textbf{\bibinfo {volume} {54}},\ \bibinfo {pages} {1392} (\bibinfo {year}
  {1985})%
  \bibAnnoteFile{NoStop}{wooten-1392}%
\bibitem{moss-775}%
  \BibitemOpen
  \bibfield{author}{%
  \bibinfo {author} {\bibfnamefont{T.~S.}\ \bibnamefont{Moss}},\ }%
  \bibfield{journal}{%
  \bibinfo {journal} {Proc. Phys. Soc. London Sect. B}\ }%
  \textbf{\bibinfo {volume} {67}},\ \bibinfo {pages} {775} (\bibinfo {year}
  {1954})%
  \bibAnnoteFile{NoStop}{moss-775}%
\bibitem{burstein-632}%
  \BibitemOpen
  \bibfield{author}{%
  \bibinfo {author} {\bibfnamefont{E.}~\bibnamefont{Burstein}},\ }%
  \bibfield{journal}{%
  \bibinfo {journal} {Phys. Rev.}\ }%
  \textbf{\bibinfo {volume} {93}},\ \bibinfo {pages} {632} (\bibinfo {year}
  {1954})%
  \bibAnnoteFile{NoStop}{burstein-632}%
\bibitem{kroger-1974}%
  \BibitemOpen
  \bibfield{author}{%
  \bibinfo {author} {\bibfnamefont{F.~A.}\
  \bibnamefont{Kr$\ddot{\mathrm{o}}$ger}},\ }%
  \emph{\bibinfo {title} {The Chemistry of Imperfect Crystals: Volume 2}},\
  \bibinfo {edition} {2nd}\ ed.\ (\bibinfo {publisher} {North-Holland},\
  \bibinfo {address} {Amsterdam},\ \bibinfo {year} {1974})%
  \bibAnnoteFile{NoStop}{kroger-1974}%
\bibitem{smith-2000}%
  \BibitemOpen
  \bibfield{author}{%
  \bibinfo {author} {\bibfnamefont{D.~M.}\ \bibnamefont{Smyth}},\ }%
  \emph{\bibinfo {title} {The defect chemistry of metal oxides}}\ (\bibinfo
  {publisher} {Oxford University Press},\ \bibinfo {address} {Oxford},\
  \bibinfo {year} {2000})%
  \bibAnnoteFile{NoStop}{smith-2000}%
\bibitem{yu-05}%
  \BibitemOpen
  \bibfield{author}{%
  \bibinfo {author} {\bibfnamefont{P.~Y.}\ \bibnamefont{Yu}}\ and\ \bibinfo
  {author} {\bibfnamefont{M.}~\bibnamefont{Cardona}},\ }%
  \emph{\bibinfo {title} {Fundamentals of Semiconductors}},\ \bibinfo {edition}
  {3rd}\ ed.\ (\bibinfo {publisher} {Springer},\ \bibinfo {address} {Berlin},\
  \bibinfo {year} {2005})%
  \bibAnnoteFile{NoStop}{yu-05}%
\bibitem{walukiewicz-123}%
  \BibitemOpen
  \bibfield{author}{%
  \bibinfo {author} {\bibfnamefont{W.}~\bibnamefont{Walukiewicz}},\ }%
  \bibfield{journal}{%
  \bibinfo {journal} {Phys. B: Cond. Mat.}\ }%
  \textbf{\bibinfo {volume} {302-303}},\ \bibinfo {pages} {123} (\bibinfo
  {year} {2001})%
  \bibAnnoteFile{NoStop}{walukiewicz-123}%
\bibitem{zhang-1232}%
  \BibitemOpen
  \bibfield{author}{%
  \bibinfo {author} {\bibfnamefont{S.~B.}\ \bibnamefont{Zhang}}, \bibinfo
  {author} {\bibfnamefont{S.-H.}\ \bibnamefont{Wei}},\ and\ \bibinfo {author}
  {\bibfnamefont{A.}~\bibnamefont{Zunger}},\ }%
  \bibfield{journal}{%
  \Doi{10.1103/PhysRevLett.84.1232}{\bibinfo {journal} {Phys. Rev. Lett.}}\ }%
  \textbf{\bibinfo {volume} {84}},\ \bibinfo {pages} {1232} (\bibinfo {year}
  {2000})%
  \bibAnnoteFile{NoStop}{zhang-1232}%
\bibitem{xu-543}%
  \BibitemOpen
  \bibfield{author}{%
  \bibinfo {author} {\bibfnamefont{Y.}~\bibnamefont{Xu}}\ and\ \bibinfo
  {author} {\bibfnamefont{M.~A.}\ \bibnamefont{Schoonenz}},\ }%
  \bibfield{journal}{%
  \bibinfo {journal} {Am. Mineral.}\ }%
  \textbf{\bibinfo {volume} {85}},\ \bibinfo {pages} {543} (\bibinfo {year}
  {2000})%
  \bibAnnoteFile{NoStop}{xu-543}%
\bibitem{kowalczyk-1605}%
  \BibitemOpen
  \bibfield{author}{%
  \bibinfo {author} {\bibfnamefont{S.~P.}\ \bibnamefont{Kowalczyk}}, \bibinfo
  {author} {\bibfnamefont{J.~T.}\ \bibnamefont{Cheung}}, \bibinfo {author}
  {\bibfnamefont{E.~A.}\ \bibnamefont{Kraut}},\ and\ \bibinfo {author}
  {\bibfnamefont{R.~W.}\ \bibnamefont{Grant}},\ }%
  \bibfield{journal}{%
  \bibinfo {journal} {Phys. Rev. Lett.}\ }%
  \textbf{\bibinfo {volume} {56}},\ \bibinfo {pages} {1605} (\bibinfo {year}
  {1986})%
  \bibAnnoteFile{NoStop}{kowalczyk-1605}%
\bibitem{wei-2011}%
  \BibitemOpen
  \bibfield{author}{%
  \bibinfo {author} {\bibfnamefont{S.-H.}\ \bibnamefont{Wei}}\ and\ \bibinfo
  {author} {\bibfnamefont{A.}~\bibnamefont{Zunger}},\ }%
  \bibfield{journal}{%
  \bibinfo {journal} {Appl. Phys. Lett.}\ }%
  \textbf{\bibinfo {volume} {72}},\ \bibinfo {pages} {2011} (\bibinfo {year}
  {1998})%
  \bibAnnoteFile{NoStop}{wei-2011}%
\bibitem{wei-144}%
  \BibitemOpen
  \bibfield{author}{%
  \bibinfo {author} {\bibfnamefont{S.-H.}\ \bibnamefont{Wei}}\ and\ \bibinfo
  {author} {\bibfnamefont{A.}~\bibnamefont{Zunger}},\ }%
  \bibfield{journal}{%
  \Doi{10.1103/PhysRevLett.59.144}{\bibinfo {journal} {Phys. Rev. Lett.}}\ }%
  \textbf{\bibinfo {volume} {59}},\ \bibinfo {pages} {144} (\bibinfo {year}
  {1987})%
  \bibAnnoteFile{NoStop}{wei-144}%
\bibitem{li-212109}%
  \BibitemOpen
  \bibfield{author}{%
  \bibinfo {author} {\bibfnamefont{Y.-H.}\ \bibnamefont{Li}}, \bibinfo {author}
  {\bibfnamefont{A.}~\bibnamefont{Walsh}}, \bibinfo {author}
  {\bibfnamefont{S.}~\bibnamefont{Chen}}, \bibinfo {author}
  {\bibfnamefont{W.-J.}\ \bibnamefont{Yin}}, \bibinfo {author}
  {\bibfnamefont{J.-H.}\ \bibnamefont{Yang}}, \bibinfo {author}
  {\bibfnamefont{J.}~\bibnamefont{Li}}, \bibinfo {author} {\bibfnamefont{J.~L.
  F.~D.}\ \bibnamefont{Silva}}, \bibinfo {author} {\bibfnamefont{X.~G.}\
  \bibnamefont{Gong}},\ and\ \bibinfo {author} {\bibfnamefont{S.-H.}\
  \bibnamefont{Wei}},\ }%
  \bibfield{journal}{%
  \Doi{10.1063/1.3143626}{\bibinfo {journal} {Appl. Phys. Lett.}}\ }%
  \textbf{\bibinfo {volume} {94}},\ \bibinfo {pages} {212109} (\bibinfo {year}
  {2009})%
  \bibAnnoteFile{NoStop}{li-212109}%
\bibitem{kamiya-1061}%
  \BibitemOpen
  \bibfield{author}{%
  \bibinfo {author} {\bibfnamefont{T.}~\bibnamefont{Kamiya}}\ and\ \bibinfo
  {author} {\bibfnamefont{M.}~\bibnamefont{Kawasaki}},\ }%
  \bibfield{journal}{%
  \bibinfo {journal} {MRS Bull.}\ }%
  \textbf{\bibinfo {volume} {33}},\ \bibinfo {pages} {1061} (\bibinfo {year}
  {2008})%
  \bibAnnoteFile{NoStop}{kamiya-1061}%
\bibitem{schleife-012104}%
  \BibitemOpen
  \bibfield{author}{%
  \bibinfo {author} {\bibfnamefont{A.}~\bibnamefont{Schleife}}, \bibinfo
  {author} {\bibfnamefont{F.}~\bibnamefont{Fuchs}}, \bibinfo {author}
  {\bibfnamefont{C.}~\bibnamefont{R\"{o}dl}}, \bibinfo {author}
  {\bibfnamefont{J.}~\bibnamefont{Furthm\"{u}ller}},\ and\ \bibinfo {author}
  {\bibfnamefont{F.}~\bibnamefont{Bechstedt}},\ }%
  \bibfield{journal}{%
  \Doi{10.1063/1.3059569}{\bibinfo {journal} {Appl. Phys. Lett.}}\ }%
  \textbf{\bibinfo {volume} {94}},\ \bibinfo {eid} {012104} (\bibinfo {year}
  {2009})%
  \bibAnnoteFile{NoStop}{schleife-012104}%
\bibitem{king-jpcm}%
  \BibitemOpen
  \bibfield{author}{%
  \bibinfo {author} {\bibfnamefont{P.~D.~C.}\ \bibnamefont{King}},\ }%
  \bibfield{journal}{%
  \bibinfo {journal} {J. Phys: Condens. Matter}\ }%
  \textbf{\bibinfo {volume} {This Issue}},\ \bibinfo {pages} {C} (\bibinfo
  {year} {2011})%
  \bibAnnoteFile{NoStop}{king-jpcm}%
\bibitem{catlow-nm}%
  \BibitemOpen
  \bibfield{author}{%
  \bibinfo {author} {\bibfnamefont{C.~R.~A.}\ \bibnamefont{Catlow}}, \bibinfo
  {author} {\bibfnamefont{A.~A.}\ \bibnamefont{Sokol}},\ and\ \bibinfo {author}
  {\bibfnamefont{A.}~\bibnamefont{Walsh}},\ }%
  \bibfield{journal}{%
  \bibinfo {journal} {In Preparation}}%
   (\bibinfo {year} {2011})%
  \bibAnnoteFile{NoStop}{catlow-nm}%
\bibitem{wei-337}%
  \BibitemOpen
  \bibfield{author}{%
  \bibinfo {author} {\bibfnamefont{S.-H.}\ \bibnamefont{Wei}},\ }%
  \bibfield{journal}{%
  \bibinfo {journal} {Comput. Mater. Sci.}\ }%
  \textbf{\bibinfo {volume} {30}},\ \bibinfo {pages} {337} (\bibinfo {year}
  {2004})%
  \bibAnnoteFile{NoStop}{wei-337}%
\bibitem{zhu-245209}%
  \BibitemOpen
  \bibfield{author}{%
  \bibinfo {author} {\bibfnamefont{Y.~Z.}\ \bibnamefont{Zhu}}, \bibinfo
  {author} {\bibfnamefont{G.~D.}\ \bibnamefont{Chen}}, \bibinfo {author}
  {\bibfnamefont{H.}~\bibnamefont{Ye}}, \bibinfo {author}
  {\bibfnamefont{A.}~\bibnamefont{Walsh}}, \bibinfo {author}
  {\bibfnamefont{C.~Y.}\ \bibnamefont{Moon}},\ and\ \bibinfo {author}
  {\bibfnamefont{S.-H.}\ \bibnamefont{Wei}},\ }%
  \bibfield{journal}{%
  \bibinfo {journal} {Phys. Rev. B}\ }%
  \textbf{\bibinfo {volume} {77}},\ \bibinfo {pages} {245209} (\bibinfo {year}
  {2008})%
  \bibAnnoteFile{NoStop}{zhu-245209}%
\bibitem{walsh-073105}%
  \BibitemOpen
  \bibfield{author}{%
  \bibinfo {author} {\bibfnamefont{A.}~\bibnamefont{Walsh}}, \bibinfo {author}
  {\bibfnamefont{J.~L.~F.}\ \bibnamefont{DaSilva}}, \bibinfo {author}
  {\bibfnamefont{Y.}~\bibnamefont{Yan}}, \bibinfo {author}
  {\bibfnamefont{M.~M.}\ \bibnamefont{Al-Jassim}},\ and\ \bibinfo {author}
  {\bibfnamefont{S.-H.}\ \bibnamefont{Wei}},\ }%
  \bibfield{journal}{%
  \bibinfo {journal} {Phys. Rev. B}\ }%
  \textbf{\bibinfo {volume} {7}},\ \bibinfo {pages} {073105} (\bibinfo {year}
  {2009})%
  \bibAnnoteFile{NoStop}{walsh-073105}%
\bibitem{medvedeva-57004}%
  \BibitemOpen
  \bibfield{author}{%
  \bibinfo {author} {\bibfnamefont{J.~E.}\ \bibnamefont{Medvedeva}},\ }%
  \bibfield{journal}{%
  \bibinfo {journal} {Euro. Phys. Lett.}\ }%
  \textbf{\bibinfo {volume} {78}},\ \bibinfo {pages} {57004} (\bibinfo {year}
  {2007})%
  \bibAnnoteFile{NoStop}{medvedeva-57004}%
\bibitem{pauling-1010}%
  \BibitemOpen
  \bibfield{author}{%
  \bibinfo {author} {\bibfnamefont{L.}~\bibnamefont{Pauling}},\ }%
  \bibfield{journal}{%
  \bibinfo {journal} {J. Am. Chem. Soc.}\ }%
  \textbf{\bibinfo {volume} {51}},\ \bibinfo {pages} {1010} (\bibinfo {year}
  {1929})%
  \bibAnnoteFile{NoStop}{pauling-1010}%
\bibitem{pauling-1960}%
  \BibitemOpen
  \bibfield{author}{%
  \bibinfo {author} {\bibfnamefont{L.}~\bibnamefont{Pauling}},\ }%
  \emph{\bibinfo {title} {The nature of the chemical bond}},\ \bibinfo
  {edition} {3rd}\ ed.\ (\bibinfo {publisher} {Cornell University Press},\
  \bibinfo {address} {Ithaca, New York},\ \bibinfo {year} {1960})%
  \bibAnnoteFile{NoStop}{pauling-1960}%
\bibitem{cox-2601}%
  \BibitemOpen
  \bibfield{author}{%
  \bibinfo {author} {\bibfnamefont{S.~F.~J.}\ \bibnamefont{Cox}}, \bibinfo
  {author} {\bibfnamefont{E.~A.}\ \bibnamefont{Davis}}, \bibinfo {author}
  {\bibfnamefont{S.~P.}\ \bibnamefont{Cottrell}}, \bibinfo {author}
  {\bibfnamefont{P.~J.~C.}\ \bibnamefont{King}}, \bibinfo {author}
  {\bibfnamefont{J.~S.}\ \bibnamefont{Lord}}, \bibinfo {author}
  {\bibfnamefont{J.~M.}\ \bibnamefont{Gil}}, \bibinfo {author}
  {\bibfnamefont{H.~V.}\ \bibnamefont{Alberto}}, \bibinfo {author}
  {\bibfnamefont{R.~C.}\ \bibnamefont{Vil\~ao}}, \bibinfo {author}
  {\bibfnamefont{J.}~\bibnamefont{Piroto~Duarte}}, \bibinfo {author}
  {\bibfnamefont{N.}~\bibnamefont{Ayres~de Campos}}, \bibinfo {author}
  {\bibfnamefont{A.}~\bibnamefont{Weidinger}}, \bibinfo {author}
  {\bibfnamefont{R.~L.}\ \bibnamefont{Lichti}},\ and\ \bibinfo {author}
  {\bibfnamefont{S.~J.~C.}\ \bibnamefont{Irvine}},\ }%
  \bibfield{journal}{%
  \Doi{10.1103/PhysRevLett.86.2601}{\bibinfo {journal} {Phys. Rev. Lett.}}\ }%
  \textbf{\bibinfo {volume} {86}},\ \bibinfo {pages} {2601} (\bibinfo {year}
  {2001})%
  \bibAnnoteFile{NoStop}{cox-2601}%
\bibitem{king-081201}%
  \BibitemOpen
  \bibfield{author}{%
  \bibinfo {author} {\bibfnamefont{P.~D.~C.}\ \bibnamefont{King}}, \bibinfo
  {author} {\bibfnamefont{R.~L.}\ \bibnamefont{Lichti}}, \bibinfo {author}
  {\bibfnamefont{Y.~G.}\ \bibnamefont{Celebi}}, \bibinfo {author}
  {\bibfnamefont{J.~M.}\ \bibnamefont{Gil}}, \bibinfo {author}
  {\bibfnamefont{R.~C.}\ \bibnamefont{Vil\~ao}}, \bibinfo {author}
  {\bibfnamefont{H.~V.}\ \bibnamefont{Alberto}}, \bibinfo {author}
  {\bibfnamefont{J.}~\bibnamefont{Piroto~Duarte}}, \bibinfo {author}
  {\bibfnamefont{D.~J.}\ \bibnamefont{Payne}}, \bibinfo {author}
  {\bibfnamefont{R.~G.}\ \bibnamefont{Egdell}}, \bibinfo {author}
  {\bibfnamefont{I.}~\bibnamefont{McKenzie}}, \bibinfo {author}
  {\bibfnamefont{C.~F.}\ \bibnamefont{McConville}}, \bibinfo {author}
  {\bibfnamefont{S.~F.~J.}\ \bibnamefont{Cox}},\ and\ \bibinfo {author}
  {\bibfnamefont{T.~D.}\ \bibnamefont{Veal}},\ }%
  \bibfield{journal}{%
  \Doi{10.1103/PhysRevB.80.081201}{\bibinfo {journal} {Phys. Rev. B}}\ }%
  \textbf{\bibinfo {volume} {80}},\ \bibinfo {pages} {081201} (\bibinfo {year}
  {2009})%
  \bibAnnoteFile{NoStop}{king-081201}%
\bibitem{king-062110}%
  \BibitemOpen
  \bibfield{author}{%
  \bibinfo {author} {\bibfnamefont{P.~D.~C.}\ \bibnamefont{King}}, \bibinfo
  {author} {\bibfnamefont{I.}~\bibnamefont{McKenzie}},\ and\ \bibinfo {author}
  {\bibfnamefont{T.~D.}\ \bibnamefont{Veal}},\ }%
  \bibfield{journal}{%
  \bibinfo {journal} {Appl. Phys. Lett.}\ }%
  \textbf{\bibinfo {volume} {96}} (\bibinfo {year} {2010})%
  \bibAnnoteFile{NoStop}{king-062110}%
\bibitem{walsh-547}%
  \BibitemOpen
  \bibfield{author}{%
  \bibinfo {author} {\bibfnamefont{A.}~\bibnamefont{Walsh}}, \bibinfo {author}
  {\bibfnamefont{Y.}~\bibnamefont{Yan}}, \bibinfo {author}
  {\bibfnamefont{M.~N.}\ \bibnamefont{Huda}}, \bibinfo {author}
  {\bibfnamefont{M.~M.}\ \bibnamefont{Al-Jassim}},\ and\ \bibinfo {author}
  {\bibfnamefont{S.-H.}\ \bibnamefont{Wei}},\ }%
  \bibfield{journal}{%
  \Doi{10.1021/cm802894z}{\bibinfo {journal} {Chem. Mater.}}\ }%
  \textbf{\bibinfo {volume} {21}},\ \bibinfo {pages} {547} (\bibinfo {year}
  {2009})%
  \bibAnnoteFile{NoStop}{walsh-547}%
\bibitem{ogo-032113}%
  \BibitemOpen
  \bibfield{author}{%
  \bibinfo {author} {\bibfnamefont{Y.}~\bibnamefont{Ogo}}, \bibinfo {author}
  {\bibfnamefont{H.}~\bibnamefont{Hiramatsu}}, \bibinfo {author}
  {\bibfnamefont{K.}~\bibnamefont{Nomura}}, \bibinfo {author}
  {\bibfnamefont{H.}~\bibnamefont{Yanagi}}, \bibinfo {author}
  {\bibfnamefont{T.}~\bibnamefont{Kamiya}}, \bibinfo {author}
  {\bibfnamefont{M.}~\bibnamefont{Hirano}},\ and\ \bibinfo {author}
  {\bibfnamefont{H.}~\bibnamefont{Hosono}},\ }%
  \bibfield{journal}{%
  \Doi{10.1063/1.2964197}{\bibinfo {journal} {Appl. Phys. Lett.}}\ }%
  \textbf{\bibinfo {volume} {93}},\ \bibinfo {eid} {032113} (\bibinfo {year}
  {2008})%
  \bibAnnoteFile{NoStop}{ogo-032113}%
\bibitem{vinke-83}%
  \BibitemOpen
  \bibfield{author}{%
  \bibinfo {author} {\bibfnamefont{I.}~\bibnamefont{Vinke}}, \bibinfo {author}
  {\bibfnamefont{J.}~\bibnamefont{Diepgrond}}, \bibinfo {author}
  {\bibfnamefont{B.}~\bibnamefont{Boukamp}}, \bibinfo {author}
  {\bibfnamefont{K.}~\bibnamefont{de~Vries}},\ and\ \bibinfo {author}
  {\bibfnamefont{A.}~\bibnamefont{Burggraaf}},\ }%
  \bibfield{journal}{%
  \bibinfo {journal} {Solid State Ionics}\ }%
  \textbf{\bibinfo {volume} {6}},\ \bibinfo {pages} {83} (\bibinfo {year}
  {1992})%
  \bibAnnoteFile{NoStop}{vinke-83}%
\bibitem{cheetham-58}%
  \BibitemOpen
  \bibfield{author}{%
  \bibinfo {author} {\bibfnamefont{A.~K.}\ \bibnamefont{Cheetham}}\ and\
  \bibinfo {author} {\bibfnamefont{C.~N.~R.}\ \bibnamefont{Rao}},\ }%
  \bibfield{journal}{%
  \Doi{10.1126/science.1147231}{\bibinfo {journal} {Science}}\ }%
  \textbf{\bibinfo {volume} {318}},\ \bibinfo {pages} {58} (\bibinfo {year}
  {2007})%
  \bibAnnoteFile{NoStop}{cheetham-58}%
\bibitem{walsh-1284}%
  \BibitemOpen
  \bibfield{author}{%
  \bibinfo {author} {\bibfnamefont{A.}~\bibnamefont{Walsh}},\ }%
  \bibfield{journal}{%
  \bibinfo {journal} {J. Phys. Chem. Lett.}\ }%
  \textbf{\bibinfo {volume} {1}},\ \bibinfo {pages} {1284} (\bibinfo {year}
  {2010})%
  \bibAnnoteFile{NoStop}{walsh-1284}%
\bibitem{walsh-2341}%
  \BibitemOpen
  \bibfield{author}{%
  \bibinfo {author} {\bibfnamefont{A.}~\bibnamefont{Walsh}}\ and\ \bibinfo
  {author} {\bibfnamefont{C.~R.~A.}\ \bibnamefont{Catlow}},\ }%
  \bibfield{journal}{%
  \bibinfo {journal} {ChemPhysChem}\ }%
  \textbf{\bibinfo {volume} {11}},\ \bibinfo {pages} {2341} (\bibinfo {year}
  {2010})%
  \bibAnnoteFile{NoStop}{walsh-2341}%
\end{thebibliography}

%

\end{document}